\documentclass[useAMS,usenatbib,usegraphicx]{pasj00}
\RelativeFigurePath{figs}
\begin{document}

\SetRunningHead{A.~T.~Okazaki et al.}
               {Origin of Two Types of X-Ray Outbursts in Be/X-Ray Binaries. I.~Accretion Scenarios}
\Received{2012/09/15}%
\Accepted{2012/11/21}

\title{Origin of Two Types of X-Ray Outbursts in Be/X-Ray Binaries. I.~Accretion Scenarios}


\author{Atsuo T. \textsc{Okazaki}\altaffilmark{1}, 
Kimitake \textsc{Hayasaki}\altaffilmark{2,3} 
and Yuki \textsc{Moritani}\altaffilmark{2,4}} 
\altaffiltext{1}{Faculty of Engineering, Hokkai-Gakuen University, Toyohira-ku,
       Sapporo 062-8605, Japan}
\email{okazaki@lst.hokkai-s-u.ac.jp}
\altaffiltext{2}{Department of Astronomy, Kyoto University, Oiwake-cho, 
Kitashirakawa, Sakyo-ku, Kyoto 606-8502, Japan}
\altaffiltext{3}{Korea Astronomy and Space Science Institute, Daedeokdaero 776, Yuseong, Daejeon 305-348, Korea} 
\altaffiltext{4}{Hiroshima Astrophysical Science Center, Hiroshima University
1-3-1 Kagamiyama Higashi-Hiroshima City 739-8511, Japan}


\KeyWords{stars: binaries, emission-line---Be, pulsars: individual (A~0535$+$262, 4U\,0115$+$634), X-rays: binaries}

\maketitle

\begin{abstract}
We propose the new scenario for X-ray outbursts in Be/X-ray binaries that normal and giant 
outbursts are respectively caused by radiatively inefficient accretion flows (RIAFs) and Bondi-Hoyle-Lyttleton (BHL) accretion of the material transferred from the outermost part of a Be disk misaligned with the binary orbital plane. Based on simulated mass-transfer rates from misaligned Be disks,
together with simplified accretion flow models, 
we show that mass-accretion rates estimated from the luminosity of the normal X-ray outbursts are consistent with those obtained with advection-dominated accretion flows, not with the 
standard, radiative-cooling dominated, accretion.
Our RIAF scenario for normal X-ray outbursts resolves problems that have challenged the standard disk picture for these outbursts. When a misaligned Be disk 
crosses the orbit of the neutron star, e.g., by warping, 
the neutron star can capture a large amount of mass 
via BHL-type accretion during the disk transit event. We numerically show that 
such a process can reproduce the X-ray luminosity of giant X-ray outbursts.
In the case of very high Be disk density, 
the accretion flow associated with the disk transit becomes supercritical, 
giving rise to the luminosity higher than the Eddington luminosity.
\end{abstract}

\section{Introduction}
\label{sec:intro}

%
Be/X-ray binaries are high-mass X-ray sources comprised of 
a Be star and a neutron star in a wide and, in most cases, eccentric orbit.
A Be star is a non-supergiant early-type star with a circumstellar disk
formed by viscous diffusion of matter ejected from the star.
This \textit{decretion} disk is geometrically thin and Keplerian, 
as are viscous accretion disks \citep{Lee1991},
and is, in normal condition, truncated by 
the tidal/resonant interaction with the neutron star (e.g., \cite{Okazaki2001}).
Because of its eccentric orbit, the neutron star can capture gas from the Be disk 
only for a short span of time during the close encounter with the Be disk.
Such a brief interaction makes the Be/X-ray binaries transient X-ray sources.
They show two types of X-ray outbursts: normal (Type I) outbursts, of which 
the X-ray luminosity $L_\mathrm{X} \sim 10^{36-37}\,\mathrm{erg\;s}^{-1}$ 
and the interval is the orbital period, and giant (Type II) outbursts, which are 
significantly brighter ($L_\mathrm{X} > 10^{37}\,\mathrm{erg\;s}^{-1}$) and 
less frequent than normal outbursts \citep{Stella1986, Negueruela1998}. 

%
Thanks to long-term, multi-wavelength monitoring observations of Be/X-ray 
binaries, it is now widely accepted that normal outbursts are triggered by 
the mass transfer from a tidally truncated Be disk at or near periastron 
passage and that giant outbursts are somehow associated with warping 
episodes of the Be disk. However, the details of the interaction between the 
Be disk and the neutron star and resultant accretion flows remain elusive.

%
In previous studies, the accretion disk which forms around the neutron star has 
been implicitly assumed to be modeled by the standard disk 
\citep{Shakura1973}. It is geometrically thin and
Keplerian, supported by rotation in the radial direction and 
by gas pressure in the vertical direction, in which the heating by viscosity is balanced with
the cooling by radiation from the disk surface at each radius
(e.g., \cite{Frank2002,Kato2008}).
Several observed features, however, challenge the standard disk picture for Be/X-ray binaries.
For instance, the spin-up episode of the X-ray pulsar in Be/X-ray binaries are, 
in general, seen only during X-ray outbursts (e.g., \cite{Bildsten1997} for A~0535$+$262 
and GRO~J1744$+$28; \cite{Parmar1989, Wilson2008} for EXO~2030$+$375), 
which often lasts for a much shorter time than the orbital period.
This suggests that the accretion disk, if formed, is transient. 
Large and rapid flux variations in X-ray outbursts 
also support the idea of transient accretion disks.
In contrast, the standard disk theory suggests that the accretion disk is not transient 
but persistent, because the viscous- accretion timescale is generally 
much longer than the orbital period.

It also remains unclear what mechanism triggers giant X-ray outbursts.    
There is observational evidence for warping of the outer part of the 
Be disk before/during giant X-ray outbursts. This episode 
has most clearly been seen in 4U~0115$+$634 \citep{Negueruela2001b, Reig2007}, 
where the H$\alpha$ emission line profile, which was usually double-peaked, varied 
between a single peaked profile and a profile with absorption core, with the timescale 
of a year or so. Although single peaked profiles and those with absorption core are usually 
observed for Be stars seen from the polar and the equatorial directions, respectively, 
the spin axis of the Be star cannot vary with such a short timescale. Thus, this drastic 
line-profile variation indicates that the angle between the observer and 
the Be disk plane varied according to the precession of a warped Be disk 
\citep{Negueruela2001b,Reig2007}.

4U~0115$+$634 is not the only system that has shown the warping episode of 
the Be disk. Recent spectroscopic monitoring observations of another Be/X-ray 
binary A~0535$+$262 have detected very complicated changes in the H$\alpha$ 
line profiles during and after the 2009 giant X-ray outburst \citep{Moritani2011}.
Analyzing the variability of several line profiles, \citet{Moritani2012} found that 
the observed features can be explained by a precessing, warped Be disk, which 
is misaligned with the binary orbital plane.

We note that the association of the warping episode with giant X-ray outbursts 
favors systems with misaligned Be disks 
over those with Be disks coplanar with the orbital plane.
In misaligned systems, the warping can bend the disk outer part towards 
the orbital plane, enabling the neutron star to capture more material from the 
Be disk than it does without warping. On the other hand, in aligned systems 
where the Be disks are coplanar with the binary orbital plane, it is difficult to 
understand how the warping of initially coplanar disks can enhance the mass 
supply to the neutron star. Given that the eccentric orbit of Be/X-ray binaries is  
attributed to a supernova kick, the same mechanism is also likely to give rise 
to the misalignment between the Be star's spin axis and the orbital axis of the 
binary. Therefore, it is important to study dynamical interactions in 
misaligned systems, despite that there is little compelling evidence of the 
misalignment between the Be disk and the binary orbital plane in Be/X-ray 
binaries.

In this paper, we examine accretion processes in Be/X-ray binaries, 
with particular attention to A~0535$+$262 and 4U~0115$+$634, 
two best-studied systems with rather different orbital parameters.
Based on the study of these systems, we propose that normal 
and giant X-ray outbursts in Be/X-ray binaries are respectively caused 
by radiatively inefficient accretion flows (RIAFs) and 
Bondi-Hoyle-Lyttleton (BHL) accretion of the material transferred 
from a Be disk misaligned with the binary orbital plane,
and that giant outbursts occur
when the tilted Be disk crosses
the orbit of the neutron star in the direction of periastron.

The paper is organized as follows: We first argue, in section~\ref{sec:sadm}, 
that the standard disk accretion picture does not work as a model for X-ray outbursts of Be/X-ray binaries. 
Then, in section~\ref{sec:type1}, we show that RIAFs
naturally explain characteristic features of normal X-ray outbursts.
In section~\ref{sec:type2}, we point out that the accretion rate and timescale 
evaluated on the basis of our scenario for giant X-ray outbursts are consistent with 
those observed.
We also discuss in this section the possibility of supercritical accretion flows 
onto the neutron star in Be/X-ray binaries.
Section~\ref{sec:discussion} is devoted for discussion about observational implications
of our scenarios, and section~\ref{sec:conclusions} concludes the paper.

%
\section{Classical Picture for X-ray Outbursts}
\label{sec:sadm}
%

As summarized in the next section, the normal X-ray outbursts last only 
for a small fraction of the orbital period. This short timescale challenges 
the classical, standard disk picture for Be/X-ray binaries. In this section, we first show 
that the standard disk model has too long accretion timescale 
to explain the duration time of normal X-ray outbursts. Afterward, we 
discuss a couple of issues, where the standard disk picture is 
inconsistent with observations. We present alternative scenarios for 
X-ray outbursts in later sections.

%
\subsection{Viscous Timescale}
\label{subsec:tvis}
%
In the standard disk, accretion occurs in the viscous timescale 
$t_\mathrm{vis}$, which is given by
\begin{equation}
t_\mathrm{vis}=\alpha^{-1} \left( \frac{r}{H} \right)^2
 \frac{1}{\Omega_\mathrm{K}}
\label{eq:tvis}
\end{equation}
(e.g., \cite{Frank2002,Kato2008}), 
where $\alpha$ is the Shakura-Sunyaev's viscosity parameter, 
$r$ is the radius measured from the neutron star, $H$ 
is the vertical scale-height of the accretion disk at $r$,
and $\Omega_\mathrm{K}=(GM_X/r^3)^{1/2}$ with 
$M_X$ being the mass of the neutron star.
A steady $\alpha$-disk solution has the scale-height given by
\begin{equation}
\frac{H}{r} = 5.6 \times 10^{-3} \alpha_{0.1}^{-1/10} \dot{m}_{16}^{3/20} m_{1.4}^{-1/4}
   \hat{r}^{1/8},
\label{eq:hr}
\end{equation}
where $\alpha_{0.1}=\alpha/0.1$, $\dot{m}_{16}=\dot{M}/10^{16}\,\mathrm{g\;s}^{-1}$, 
$m_{1.4}=M_\mathrm{X}/1.4M_{\odot}$, 
and $\hat{r}$ is the radius normalized by the Schwarzschild radius 
$r_{\rm S}=2GM_X/c^2\approx4.1\times10^{5}m_{1.4}\,\,\rm{cm}$.
Substituting equation~(\ref{eq:hr}) into equation~(\ref{eq:tvis}), we have
\begin{equation}
t_\mathrm{vis} = 6.2 \alpha_{0.1}^{-4/5} 
   \dot{m}_{16}^{-3/10} m_{1.4}^{3/2} \hat{r}^{5/4}
   \,\,\rm{s}.
\label{eq:tvis2}
\end{equation}
In order to evaluate equation~(\ref{eq:tvis2}), we need to estimate 
the outer radius of the accretion disk and the mass-accretion rate there.
In the next subsection, we estimate $t_\mathrm{vis}$ by utilizing results 
from numerical simulations 
of tidal interaction in the Be/X-ray binaries A~0535$+$262 and 4U~0115$+$634.

\begin{table*}
\caption{System parameters for A~0535$+$262 and 4U 0115$+$634}
\begin{center}
 \begin{tabular}{p{0.01\textwidth}p{0.28\textwidth}cc}
 \hline
 && A~0535$+$262 & 4U 0115$+$634 \\ 
 \hline
 \multicolumn{4}{l}{Be star parameters} \\
 & Name	& V725 Tau	& V635 Cas \\
 & Spectral type	& O9.7IIIe \footnotemark[1]	& B0.2Ve \footnotemark[2] \\
 & Mass $M_{*}$ ($M_{\odot}$) & 25 \footnotemark[3] & 19 \footnotemark[3] \\
 & Radius $R_{*}$ ($R_{\odot}$) & 15 \footnotemark[3] & 8 \footnotemark[3] \\ 
 & Effective temperature (K) & 26,000 & 26,000 \\
 \hline
 \multicolumn{4}{l}{Be disk parameters} \\
 & Temperature $T_{\rm d}$ (K) & 15,600 & 15,600 \\
 \hline
 \multicolumn{4}{l}{Neutron star parameters} \\
 & Mass $M_X$ ($M_{\odot}$) & 1.4 & 1.4	\\
 & Radius $R_X$ (cm)  & $10^{6}$ & $10^{6}$ \\
 & Spin period $P_\mathrm{s}$ (s) & 103 \footnotemark[4] & 3.6 \footnotemark[5]	\\
 & Magnetic field strength (G) & $4.3 \times 10^{12}$ \footnotemark[6] & 0.87--1.07$\times 10^{12}$ \footnotemark[7]  \\
 \hline
 \multicolumn{4}{l}{Orbital parameters} \\
 & Orbital period $P_{\rm orb}$ (d) & 110.2 \footnotemark[8] & 24.3 \footnotemark[5]  \\
 & Orbital eccentricity $e$ & 0.47 \footnotemark[9] & 0.34 \footnotemark[5]	\\
 & Mass ratio $q$ & 0.056 & 0.074 \\
 & Semi-major axis $a$ (cm) & $2.0 \times 10^{13}$ & $6.7 \times 10^{12}$ \\
 \hline
 \multicolumn{4}{p{0.93\textwidth}}{ 
 1: \citet{Giangrande1980}, 2: \citet{Negueruela2001a}, 
 3: \citet{Vacca1996}, 
 4: \citet{Coe1975}, 5: \citet{Rappaport1978}, 
 6: \citet{Makishima1999}, 7: \citet{Ferrigno2009},
 8: \citet{Moritani2010}, 9: \citet{Finger1994}
 } \\
 \end{tabular}
\end{center}
\label{tbl:params}
\end{table*}
%
%
\subsection{Accretion Timescale Estimated from Numerical Simulations}
\label{subsec:accrate-sph}
%
%
\subsubsection{Numerical Method}
\label{sec:sph-code}
%
The simulations presented below were performed with a 3D SPH code, which is
a particle method that divides the fluid into a set of discrete \lq\lq fluid elements''
($=$particles), and is flexible in setting various initial configurations.
The code is the same as that used by \citet{Okazaki2002} (see also \cite{Bate1995}).
Using a variable smoothing length, the SPH equations with a standard
cubic-spline kernel are integrated with an individual time step for
each particle.

In our code, the Be disk is modeled by an ensemble of
gas particles with negligible self-gravity. For simplicity, the gas
particles are assumed to be isothermal at $0.6\, T_\mathrm{eff}$ with $T_\mathrm{eff}$ 
being the effective temperature of the Be star \citep{Carciofi2006}. 
We model the mass ejection process from the star by injecting gas particles 
at a constant rate just outside the equatorial surface of the star.
Once injected, gas particles interact with each other. As a result, most of the injected
particles fall back on to the Be star by losing the angular momentum and a
small fraction of particles drift outwards, obtaining the angular
momentum from the other particles.

On the other hand, the Be star and the neutron star are represented by sink
particles with the appropriate gravitational mass. Gas particles that
fall within a specified accretion radius are accreted by the sink
particle. We assume that the accretion radius of the Be star is equal to the stellar radius
$R_{*}$, while for the neutron star, we adopt a variable accretion radius of $r_{\rm L}$, 
where $r_{\rm L}$ is the Roche-lobe radius given approximately by
\begin{equation}
r_\mathrm{L} = D \frac{0.49q^{2/3}}{0.69q^{2/3} + \ln (1+q^{1/3})}
\label{eq:roche}
\end{equation}
\citep{Eggleton1983} with the mass ratio $q=M_{X}/M_{*}$, where
$M_{*}$ is the mass of the Be star,
and the instantaneous distance between the stars, $D$.

In simulations shown in this paper, we adopt artificial viscosity parameters 
$\alpha_\mathrm{SPH}=1$ or $\beta_\mathrm{SPH}=0$, which roughly corresponds to 
the Shakura-Sunyaev viscosity parameter $\alpha=0.1$ except in very-low density regions
(e.g., \cite{Okazaki2002}).
We set the binary orbit on the $x$-$y$ plane with the major axis
along the $x$-axis (the apastron is in the $+x$-direction).  At $t=0$, the neutron star is at apastron.
It orbits about the Be star primary with the orbital period $P_\mathrm{orb}$ and 
the orbital eccentricity $e$.

%
\subsubsection{A~0535$+$262}
\label{sec:sph-results-a0535}

A~0535$+$262 consists of an O9.7IIIe star and an X-ray pulsar 
with a relatively wide [$P_\mathrm{orb}=110.2\,\mathrm{d}$ \citep{Moritani2010}] 
and eccentric [$e=0.47$ \citep{Finger1994}] orbit. 
As the mass and radius of the Be star, we take $25\,M_{\odot}$ and $15\,R_{\odot}$ 
from \citet{Vacca1996}. 
With these parameters, the semi-major axis is $a = 2.0 \times 10^{13}\,\mathrm{cm}$.
The magnetic field strength of the neutron star is $4.3  \times 10^{12}\,\mathrm{G}$
from the observed cyclotron resonance scattering features 
\citep{Makishima1999}. 
Table~\ref{tbl:params} summarizes system parameters used in this paper.

We have performed two simulations for A~0535$+$262. In one simulation, the Be disk is
coplanar with the binary orbital plane, while in the other simulation,
it is misaligned with the orbital plane by $45^\circ$ about the semi-minor axis.
Each simulation has run until the Be disk is fully developed and start showing 
a regular orbital modulation in the mass capture rate by the neutron star.
The number of gas particles at the end of coplanar and misaligned disk simulations are
$\sim 170,000$ ($t=30\,P_\mathrm{orb}$) and $\sim 210,000$ ($t=50\,P_\mathrm{orb}$), respectively.

\begin{figure}[!t]
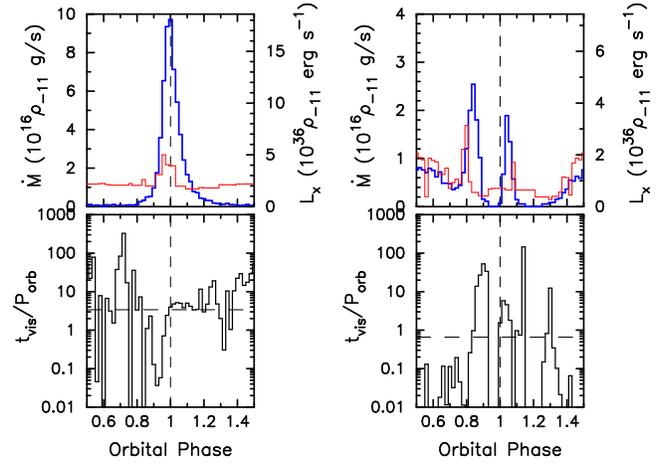

%
\centerline{
\begin{tabular}{p{0.22\textwidth}p{0.22\textwidth}}
(a) A~0535$+$262 with a coplanar Be disk & (b) A~0535$+$262 with a misaligned Be disk\\
\FigureFile(0.23\textwidth,0.23\textheight){figure1a.eps} &
\FigureFile(0.23\textwidth,0.23\textheight){figure1b.eps} \\
\end{tabular}}
 \caption{Simulation result for A~0535$+$262 with (a) a Be disk coplanar with 
 the binary orbital plane and 
 (b) a Be disk misaligned with the orbital plane by $45^\circ$ about the semi-minor axis.
 {\it Upper panels}: the orbital modulation 
 in the mass capture rate by the neutron star [thick (blue) lines] and 
 estimated mass-accretion rate at the neutron star radius [thin (red) lines]. 
 The right axis shows
 the X-ray luminosity corresponding to the mass-accretion rate.
 {\it Lower panels}: the viscous timescale at the circularization radius of captured material.
 The data is folded on the orbital period over $25 P_\mathrm{orb} \le t \le
 30 P_\mathrm{orb}$ in the coplanar case and $45 P_\mathrm{orb} \le t \le
 50 P_\mathrm{orb}$ in the misaligned case. 
 The mass-capture/accretion rate and the X-ray luminosity are measured in units of 
 $10^{16} \rho_{-11}\;\mathrm{g\;s}^{-1}$ and $10^{36} \rho_{-11}\;\mathrm{erg\;s}^{-1}$,
 respectively,
 where $\rho_{-11}$ is the base density of the Be disk
 normalized by $10^{-11}\;\mathrm{g\;cm}^{-3}$, a typical value for Be stars. 
 The dashed line in each lower panel denotes the mass-weighted average of the viscous timescale.
  }
 \label{figure1}
\end{figure}

Figure~\ref{figure1} shows the result from (a) the coplanar disk 
simulation and (b) the misaligned disk simulation. In the upper panels, the thick (blue) 
lines denote the mass capture rate by the neutron star as a function of orbital phase. 
We note that in the coplanar case, mass capture rate has a single peak at the periastron,
while it is multi-peaked when the Be disk is highly inclined from
the binary orbital plane. This is because in misaligned disks, the neutron star makes
close disk encounter twice an orbit at orbital phases different from the periastron.
We also note that the neutron star captures more mass from coplanar Be disks 
than from misaligned ones. 
This is because the distance between the neutron star
and the Be disk in the coplanar system is shorter than that for the misaligned 
Be disk rotated around the binary semi-minor axis.

In the coplanar disk simulation, the peak mass-capture rate 
is $\dot{M} \approx 10^{17} \rho_{-11}\,\mathrm{g\;s}^{-1}$,
while in the misaligned simulation it is $\dot{M} \approx 2-3 \times 
10^{16} \rho_{-11}\,\mathrm{g\;s}^{-1}$, where $\rho_{-11}$ is 
the base density of the Be disk normalized by $10^{-11} \mathrm{g\;cm}^{-3}$, 
a typical value for Be stars. As the right axis shows, these rates would produce 
the peak X-ray luminosity of $\sim 2 \times 10^{37}\rho_{-11},\mathrm{erg\;s}^{-1}$
in the coplanar case, and $4 \times 10^{36}\rho_{-11}\,\mathrm{erg\;s}^{-1}$
in the misaligned case, if the accretion time of the captured material on to 
the neutron star were negligible. Here, we have calculated the X-ray 
luminosity by $L_\mathrm{X}=GM_\mathrm{X}\dot{M}/R_\mathrm{X}$
$\approx 1.0 \times 10^{-2} \dot{m}_{16} L_\mathrm{edd}$
with $L_\mathrm{Edd}$ being the Eddington luminosity given by
\begin{equation}
L_\mathrm{Edd} = 
\frac{4\pi cGM_\mathrm{X}}{\kappa_\mathrm{es}}
\approx 1.8 \times 10^{38} m_{1.4}
\;\mathrm{erg\;s}^{-1},
\label{eq:L_Edd}
\end{equation}
where $c$ is the speed of light and $\kappa_\mathrm{es}$ 
is the opacity of the electron scattering.
We take $R_\mathrm{X}=10^{6}\,\mathrm{cm}$ as the radius of the neutron star.

%
%

Since the Roche-lobe radius $r_\mathrm{L}$ is larger than the neutron star radius
by many orders, the accretion time from this radius to the neutron star surface
could significantly reduce the peak amplitude and broaden the shape of the accretion rate.
It is thus interesting to estimate the accretion rate at the neutron star radius.
For this purpose, we first calculated the circularization radius $r_\mathrm{circ}$
from the specific angular momentum of captured material, $j$, by assuming that
$r_\mathrm{circ}$ is approximately given by $r_\mathrm{circ} = j^2/(GM_\mathrm{X})$.
We then evaluated the viscous timescale $t_\mathrm{vis}$ at $r=r_\mathrm{circ}$, 
using equation~(\ref{eq:tvis2}). In our simulations,
the mass-weighted average of $r_\mathrm{circ}$ is $r_\mathrm{circ} \sim 3 \times 10^{11}\,\mathrm{cm}$
$\sim 0.02a$ (coplanar case) and $r_\mathrm{circ} \sim 7 \times 10^{10}\,\mathrm{cm}$
$\sim 0.004a$ (misaligned case).
The lower panels of figure~\ref{figure1} shows the resultant $t_\mathrm{vis}$ 
as a function of orbital phase.
In both simulations, the viscous (i.e., accretion) timescale turns out to be longer 
than the orbital period at most phases,
in particular, $t_\mathrm{vis} \sim 5-6\,P_\mathrm{orb}$ 
at the peak mass-transfer phase.

Using the phase dependence of mass capture rate and viscous timescale, we estimated 
the accretion rate on to the neutron star as follows. 
With the mass capture rate $\dot{M}_\mathrm{cap}(t)$ at $t$,
we approximate the amount of mass captured for a short span of time 
between $t=t_0$ and $t=t_0+\Delta t$ 
by $\dot{M}_\mathrm{cap}(t_0) \Delta t$.
As a very rough approximation of the viscous accretion process, we assume that this amount of 
mass is accreted by the neutron star over $t_\mathrm{vis}$ between $t_0+t_\mathrm{vis}/2$ and 
$t_0+3 t_\mathrm{vis}/2$ at a constant rate of $\dot{M}_\mathrm{cap}(t_0) \Delta t/t_\mathrm{vis}$.
Note that in this treatment the accretion timescale is taken as being independent of the mass capture/accretion history. The method is thus expected to provide a reasonable estimate of the accretion rate, when the accretion timescale increases with orbital phase or slowly decreases with phase with the changing timescale longer than the accretion timescale itself. 
However, it might create an artificial, spiky feature in the resulting accretion rate in case the accretion timescale rapidly decreases with orbital phase, where in real systems the rapid accretion of newly captured material will certainly be prevented by an already-existing accretion disk with a longer accretion timescale.

Applying the above procedure to the phase-dependent mass capture rate and viscous timescale,
we obtained the orbital modulation of accretion rate estimated at the neutron star radius. 
The result is shown 
by the thin (red) lines in the upper panels of figure~\ref{figure1}.
As expected from the fact that the viscous timescale is much longer than the orbital period,
the obtained mass accretion rate is mostly constant and exhibits only short and small 
enhancements at phases with short accretion timescales.
These simulated features contradict with the observational characteristics of normal X-ray outbursts,
which exhibit a large and rapid change in the X-ray flux.

%
\subsubsection{4U~0115$+$634}
\label{sec:sph-results-4u0115}

\begin{figure}[!t]
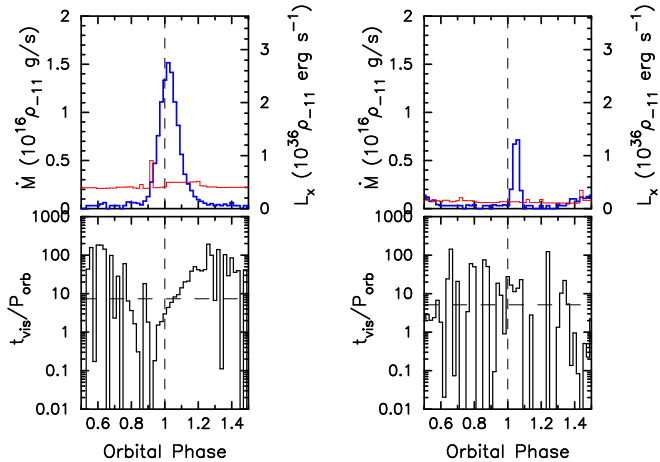

\centerline{
\begin{tabular}{p{0.23\textwidth}p{0.23\textwidth}}
(a) 4U~0115$+$634 with a coplanar Be disk & (b) 4U~0115$+$634 with a misaligned Be disk\\
\FigureFile(0.23\textwidth,0.23\textheight){figure2a.eps} &
\FigureFile(0.23\textwidth,0.23\textheight){figure2b.eps} \\
\end{tabular}}
 \caption{Simulation result for 4U~0115$+$634 with (a) a coplanar Be disk and 
 (b) a Be disk misaligned with the orbital plane by $45^\circ$ about the semi-minor axis.
 The data is folded on the orbital period over $45 P_\mathrm{orb} \le t \le
 50 P_\mathrm{orb}$ for both simulations.
 The format of the figure is the same as that of figure~\ref{figure1}.
  }
 \label{figure2}
\end{figure}
%

4U~0115$+$634 consists of a B0.2Ve star and an X-ray pulsar with a relatively narrow
($P_\mathrm{orb}=24.3\,\mathrm{d}$) and eccentric 
($e=0.34$) orbit \citep{Rappaport1978}.
As the mass and radius of the Be star, we take $19\,M_{\odot}$ and $8\,R_{\odot}$ 
from \citet{Vacca1996}.
With these parameters, the semi-major axis $a$ is 
$a = 6.7 \times 10^{12}\,\mathrm{cm}$, which is about three times
smaller than that of A~0535$+$262.
These system parameters are summarized in Table~\ref{tbl:params}.

As for A~0535$+$262, we have performed two simulations for 4U~0115$+$634, i.e.,
a coplanar disk simulation and a simulation with a Be disk misaligned by $45^\circ$ 
about the semi-minor axis.
Both simulations have run over $50 P_\mathrm{orb}$ until the Be disk is fully developed 
and start showing a regular orbital modulation in the mass capture rate by the neutron star.
The number of gas particles at the end of coplanar and misaligned simulations are
$\sim 140,000$ and $\sim 110,000$, respectively.

Figure~\ref{figure2} shows the result from (a) the coplanar 
disk simulation and (b) the misaligned disk simulation. The format of the figure 
is the same as that of figure~\ref{figure1}. From the upper panels, 
we note that the peak mass-capture rate in this system is several times lower than
in A~0535$+$262. This is because the tidal truncation of the Be disk is more 
efficient in shorter orbital period and smaller orbital eccentricity 
\citep{Okazaki2001}. Owing to a larger gap between the neutron star orbit and the 
Be disk, the neutron star of 4U~0115$+$634 has more difficulty in capturing gas 
from the Be disk than the neutron star of A~0535$+$262 does. 
By the same reason, the pre-periastron 
peak, which is caused by the pre-periastron close encounter of the neutron star to 
the Be disk, disappears in the misaligned simulation of 4U~0115$+$634. 
The stronger tidal interaction in 4U~0115$+$634 also gives rise to longer viscous 
timescale (lower panels), because the captured material has larger specific angular 
momentum. As a result, in 4U~0115$+$634 the mass-accretion rate estimated 
at the neutron star surface (red lines) are much lower and less variable than in 
A~0535$+$262
(the pre-periastron spike in Fig.~\ref{figure2}a is an artificial feature 
arising from our simplified treatment).
Apparently, these simulated features are inconsistent with the 
observational characteristics of normal X-ray outbursts.

%
\subsection{Direct Accretion vs. Propeller Regimes}
\label{subsec:acc-mode}
%

Although we have shown in the above sections that the standard disk scenario
does not agree with the observed short duration and large and rapid X-ray 
flux variations of normal X-ray outbursts, one may argue that the onset of 
propeller regime can truncate the accretion regime and hence make 
the standard disk model still viable.
Given a strong positive correlation between the spin period of the pulsar and 
the orbital period of the binary seen in Be/X-ray binaries (e.g., \cite{Corbet1986}), it is likely that
the propeller effect of the rotating magnetosphere of the pulsar plays an important role
in systems with short orbital periods. Indeed, in 4U~0115$+$634, 
\citet{Campana2001} witnessed a very rapid increase of the X-ray flux to be attributed
to the transition from the propeller regime to the direct accretion regime.
It is not clear, however, whether the same mechanism plays an important role 
in the X-ray behavior of systems with much longer orbital periods, 
such as A~0535$+$262.

Below we examine whether the propeller effect also works for A~0535$+$262,
truncating the accretion regime short and making a large difference
between the quiescent and outburst X-ray fluxes.
For this purpose, we compare the corotation radius, $r_\mathrm{C}$, where
the local Keplerian frequency is equal to the rotation frequency of
the magnetosphere, with the magnetospheric radius, $r_\mathrm{M}$,
obtained by equating the magnetic field pressure of the neutron star
to the gas pressure of the accretion disk.
For $r_\mathrm{M} < r_\mathrm{C}$, the velocity of rigidly rotating magnetic 
field is sub-Keplerian at the magnetospheric radius, which permits the material 
in the accretion disk to fall on to the neutron star along magnetic field lines. 
The system is then in the direct accretion regime. On the other hand, 
for $r_\mathrm{M} > r_\mathrm{C}$, the magnetosphere rotating at a 
super-Keplerian speed expels disk material, prohibiting the accretion to occur. 
This regime is called the propeller regime.

%
\begin{figure}[!t]
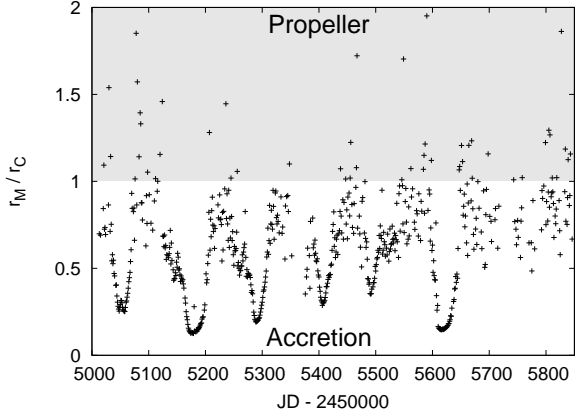

  \centering
  \FigureFile(0.45\textwidth,0.45\textwidth){figure3.eps}
    \caption{The long-term variation of the ratio of
    the magnetospheric radius $r_\mathrm{M}$ to
    the corotation radius $r_\mathrm{C}$ for A~0535$+$262. The magnetospheric radius is 
    estimated from the observed flux with \textit{Swift} with a conservative value of $k=1$.
    } 
    \label{figure3}
\end{figure}
%

The corotation radius $r_\mathrm{C}$ is given by
\begin{eqnarray}
r_\mathrm{C}=
\left(\frac{GM_XP_\mathrm{spin}^{2}}{4\pi^{2}}\right)^{1/3}
\approx 1.4 \times 10^{8} 
\,P_\mathrm{spin}^{2/3}
\,m_{1.4}^{1/3}\;
\mathrm{cm}
\label{eq:r_co}
\end{eqnarray}
where $P_{\rm spin}$ is the spin period of the neutron star.

Assuming the magnetic field of the neutron star to be
a dipole-like field, the magnetospheric radius $r_\mathrm{M}$ is written
as 
\begin{equation}
r_\mathrm{M} \approx 4.9 \times 10^8 k\,
    \dot{m}_{16}^{-2/7} m_{1.4}^{-1/7} \mu_{30}^{4/7}\;\mathrm{cm}
\label{eq:r_m}
\end{equation}
(e.g., \cite{Frank2002}).
Here, $\mu_{30}=\mu/10^{30}\,[{\rm G\,cm}^{3}]$ is the normalized magnetic moment and $k$
is a constant of order unity which depends on the physics and geometry of accretion. 
For spherical accretion, $k \sim 1$. For disk accretion, however, the previous results range
from $k \sim 0.5$ to $k \sim 1$. In this paper, we adopt $k=1$ as a conservative value for
disk accretion, which gives rise to an upper limit of  magnetospheric radius for given $\dot{m}$.
The system is in the accretion regime when the normalized accretion rate satisfies the following condition:
\begin{equation}
\dot{m}_{16} > 1.4 \times 10^{3} k^{7/2} 
P_\mathrm{spin}^{-7/3} 
m_{1.4}^{-5/3} 
\mu_{30}^2.
\label{eq:mdot-crit1}
\end{equation}
For disk accretion in A~0535$+$262, where $P_\mathrm{spin}=103\,\mathrm{s}$
and $\mu_{30}=4.3$, we have $\dot{m}_{16} \gtrsim 0.5$ for direct accretion regime.

In figure~\ref{figure3}, we compare $r_\mathrm{M}$ with $r_\mathrm{C}$ 
using the mass accretion rate estimated from the observed X-ray flux of A~0535$+$262.
We utilized the X-ray observations by $Swift$/BAT (15 -- 50 keV)
\footnote{http://swift.gsfc.nasa.gov/docs/swift/results/transients/\\1A0535p262/}, 
converting its count rates to the energy fluxes.
We assumed that 1 Crab $=$ $1.386 \times 10^{-8}$ $\mathrm{erg\;cm^{-2}\; s^{-1}}$, modeling the Crab
spectra with $n_H$ $=$ $0.26 \times 10^{22}$ $\mathrm{cm^{-2}}$, $\Gamma = 2.1$, 
and power-law normalization being 10.0.
Figure~\ref{figure3} shows the ratio of the magnetospheric radius $r_\mathrm{M}$
to the corotation radius $r_\mathrm{C}$. The figure clearly shows that
the system is almost always in the direct accretion regime, except for periods of
very low X-ray fluxes.
Therefore, we conclude that in A~0535$+$262, the large and rapid flux changes 
of normal X-ray outbursts with short duration are due to an intrinsic nature of 
the accretion flow, not controlled by the propeller effect. 
This safely rules out the standard disk scenario for normal X-ray outbursts. 

%
\section{Origin of Normal X-ray Outbursts}
\label{sec:type1}
%

The observed features of normal X-ray outbursts are summarized 
as follows: They occur at or near the periastron passage.
The X-ray luminosity increases by about one order of magnitude 
with respect to the pre-outburst state ($L_X \sim 10^{36-37}\;\mathrm{erg\;s^{-1}}$).
The duration time is a small fraction of the orbital period, typically 0.2 -- 0.3 $P_{\mathrm{orb}}$. 
Sometimes the precursors and flares occur prior to the main outbursts 
(\cite{Klochkov2010a} and \cite{Klochkov2010b} for EXO~2030$+$375; 
\cite{Caballero2008} for A~0535$+$262;
\cite{Campana2001} for 4U~0115$+$634).
Among those, the flares are interpreted to be caused by the instability of magnetosphere 
(e.g., \cite{Postnov2008}).

Although it is widely accepted that normal X-ray outbursts are triggered 
by the mass transfer from the Be disk at or near periastron,
the accretion mechanism still remains an open question.
As shown in the previous section, the standard disk scenario 
fails to explain the observational features noted above. 
Below we propose that the normal X-ray outbursts are caused 
by RIAFs on to the neutron star from a tidally truncated Be disk.

%
\subsection{Radiatively Inefficient Accretion Flows}
\label{subsec:riaf} 
%

In the context of the standard disk model, 
the viscous heating rate per unit area, $Q_+$, is assumed to 
balance with the radiative cooling rate per unit area, $Q_{\rm{rad}}$,
in the energy equation.
This approach is justified when the accretion disk is geometrically thin so that
$Q_{\rm{adv}}\sim (H/r)^2Q_+$ can be neglected, where $Q_{\rm{adv}}$ represents
the advective cooling rate per unit area.
However, in hot accretion flows with $H/r\approx c_{\rm{s}}/v_{\rm{K}} \sim 1$,
where $c_{\rm{s}}$ and $v_{\rm{K}}$ are the sound speed and Keplerian velocity of the disk,
respectively, the advection cooling cannot be neglected in the energy equation.

\citet{Abramowicz1995} found that there are two families of advection 
dominated accretion flow solutions. One family consists of optically thin, 
advection-dominated accretion flows (ADAFs),
while the other consists of optically thick, 
advection-dominated accretion flows, so-called slim disks. 
Both families are thermally and viscously stable.
From figure~3 of \citet{Abramowicz1995}, ADAF solutions exist if 
$\dot{m} \equiv \dot{M}/(L_\mathrm{Edd}/c^2) \lesssim 1$ 
(or equivalently $\dot{m}_{16} \lesssim 20 m_{1.4}$) whereas there are
slim disk solutions if $\dot{m}\gtrsim 10$ (or $\dot{m}_{16} \gtrsim 200m_{1.4}$).
In the following, we consider ADAF solutions as the RIAF model for 
normal X-ray outbursts. 
We will also discuss slim disk solutions in Section~\ref{subsec:scaf}.

%
\subsubsection{Self-similar ADAFs Solutions}
%

\citet{ny94} derived a set of solutions of 
self-similar, optically thin ADAFs with $Q_{\rm{adv}}=fQ_+$, 
where $f$ exhibits the extent to which the flow is advection dominated.
These flows give us most of important properties of  the general solutions, 
although $Q_{\rm{rad}}$ is not explicitly shown in their solutions.
Therefore, we model accretion flows onto the neutron star 
by the self-similar ADAFs when they have low accretion rates 
$\dot{m}\lesssim1$ (or $\dot{m}_{16} \lesssim 20 m_{1.4}$). 
By adopting the self-similar scaling law proposed by \citet{ny94} and \citet{ny95},
we obtain a set of solutions for $\alpha \ll 1$ as 
\begin{eqnarray}
v_{\rm{r}}
&\approx&
-2.1\times10^{9}
c_1\alpha_{0.1}\hat{r}^{-1/2} \,\,{\rm cm\;s^{-1}}, \\
H
&\approx& 
4.2\times10^{5}c_3^{1/2}m_{1.4}\hat{r}\,\,\rm{cm},\\
\rho
&\approx&
2.1\times10^{-6}c_1^{-1}c_3^{-1/2}\dot{m}_{16}\alpha_{0.1}^{-1}m_{1.4}^{-2}\hat{r}^{-3/2} 
\rm{g\,cm^{-3}},\\
T
&\approx& 
2.7\times10^{12}c_3\hat{r}^{-1}\,\,\rm{K},
\end{eqnarray}
where  $v_{\rm{r}}$ is the radial velocity, 
$\rho$ is the density, $T$ is the disk temperature, and
$c_1 \approx 0.53$, $c_2 \approx 0.34$, and $c_3 \approx 0.35$ are
numerical constants.

%
\subsubsection{Observational Implications}
%

\begin{figure}[!t]
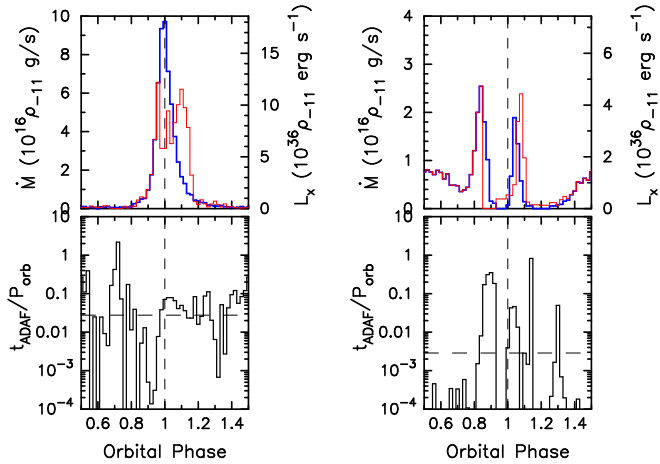

\centerline{
\begin{tabular}{p{0.23\textwidth}p{0.23\textwidth}}
(a) A~0535$+$262 with a coplanar Be disk & (b) A~0535$+$262 with a misaligned Be disk\\
\FigureFile(0.23\textwidth,0.23\textheight){figure4a.eps} &
\FigureFile(0.23\textwidth,0.23\textheight){figure4b.eps} \\
\end{tabular}}
 \caption{ADAF model for A~0535$+$262 with (a) a coplanar Be disk and 
 (b) a Be disk misaligned with the orbital plane by $45^\circ$ about the semi-minor axis.
 The format of the figure is the same as that of figure~\ref{figure1},
 except that the thin (red) lines in the upper panels denote 
 the estimated mass-accretion rate
 at the neutron star radius using the ADAF solution and that
 the lower panels show the infall timescale of 
 the ADAF at the circularization radius of captured material (solid lines)
 and the mass-weighted average of the ADAF timescale (dashed lines).}
 \label{figure4}
\end{figure}
%

\begin{figure}[!ht]
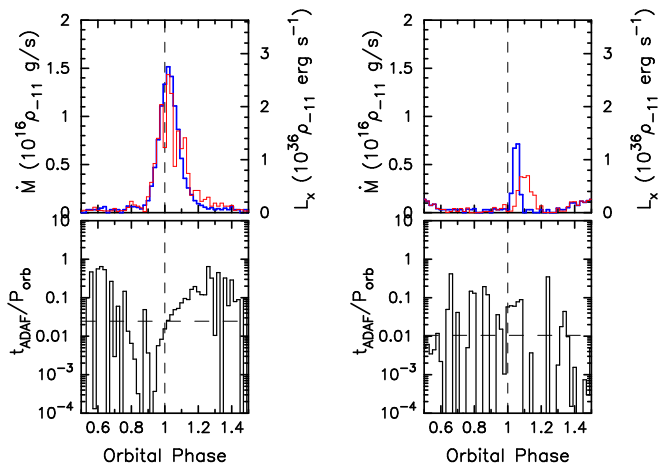

\centerline{
\begin{tabular}{p{0.23\textwidth}p{0.23\textwidth}}
(a) 4U~0115$+$634 with a coplanar Be disk & (b) 4U~0115$+$634 with a misaligned Be disk\\
\FigureFile(0.23\textwidth,0.23\textheight){figure5a.eps} &
\FigureFile(0.23\textwidth,0.23\textheight){figure5b.eps} \\
\end{tabular}}
 \caption{ADAF model for 4U~0115$+$634 with (a) a coplanar Be disk and 
 (b) a Be disk misaligned with the orbital plane by $45^\circ$ about the semi-minor axis.
 The format of the figure is the same as that of figure~\ref{figure4}.
 }
 \label{figure5}
\end{figure}
%

The infall timescale of the ADAF is evaluated by
\begin{eqnarray}
t_{\rm{ADAF}}=\frac{r}{|v_{\rm r}|}
&&
\approx 
2.0\times10^{-4}
m_{1.4}
c_1^{-1}\alpha_{0.1}^{-1}
\hat{r}^{3/2}\,\,\rm{s}.
\label{eq:t_adaf}
\end{eqnarray}
Applying equation~(\ref{eq:t_adaf}) to 
the same simulation data as in figures~\ref{figure1} and \ref{figure2} 
for A~0535$+$262 and 4U~0115$+$634, respectively, 
we have calculated the mass-accretion rate profiles in the case of ADAFs. 
The results are shown in figure~\ref{figure4} for A~0535$+$262 and 
figure~\ref{figure5} for 4U~0115$+$634. From the results for A~0535$+$262, 
we note that the ADAF solutions provide X-ray profiles with similar duration and amplitude to those of
observed normal outbursts. The pre-periastron spike of the mass-accretion rate in the coplanar simulation
is also reminiscent of an observed short spike prior to periastron (e.g., \cite{Camero-Arranz2012}). 
Comparing the mass-accretion rate in the upper panel with the $t_\mathrm{ADAF}$ plot in the lower panel,
we note that the spiky feature is caused by the material with very low specific angular momentum.

The simulated accretion rate for 4U~0115$+$634 also has the accretion timescale in agreement
with that of the observed normal outbursts. The X-ray luminosity obtained for the typical
Be disk density ($10^{-11}\,\mathrm{g\;cm}^{-3}$ at the base of the disk), however,
seems too low to be consistent with the observations. This might explain why this system
has shown normal X-ray outbursts only when the Be disk was fully developed.

Next, we evaluate the Bolometric X-ray luminosity of ADAFs.
For simplicity, we assume that the X-ray radiation emitted from the ADAF 
is produced only by the Bremsstrahlung cooling.
The cooling due to the other processes such as the 
Synchrotron radiation and Inverse Compton process will be discussed 
in a forthcoming paper in the context of high energy emissions.
The Bremsstrahlung cooling rate per unit surface area is given by
\begin{eqnarray}
Q_{\rm{br}}
&&
=\epsilon_{\rm{br}}\rho^2{T^{1/2}}H 
\nonumber \\
&&
\approx7.7\times10^{22}
c_1^{-2}
\alpha_{0.1}^{-2}
m_{1.4}^{-2}
\dot{m}_{16}
\hat{r}^{-5/2}
\,{\rm erg\,s^{-1}cm^{-2}},
\end{eqnarray}
where $\epsilon_{\rm{br}}\approx1.2\times10^{21}\,\rm{erg\,s^{-1}cm^{-2}}$ 
is the emissivity of the Bremsstrahlung radiation.
Then, the Bolometric X-ray luminosity is given by
\begin{eqnarray}
&&
L_{\rm{ADAF}}
=
2\pi{r_{\rm{S}}}^2\int_{\hat{r}_{\rm{M}}}^{\hat{r}_{\rm{circ}}}\hat{r}Q_{\rm{br}}d\hat{r}
\nonumber \\
&&
\approx
1.7\times10^{34}
\alpha_{0.1}^{-2}
k^{-1/2}
m_{1.4}^{4/7}
\mu_{30}^{-2/7}
\dot{m}_{16}^{8/7}
\,\rm{ erg\,s^{-1}}
\label{eq:ladaf}
\end{eqnarray}
for $r_{\rm M}\ll r_{\rm circ}$.

On the other hand, the X-ray luminosity $L_\mathrm{X}$ 
emitted from the surface of the neutron star is 
calculated with equation~(\ref{eq:L_Edd}) by
\begin{eqnarray}
L_\mathrm{X} = \epsilon_\mathrm{X} m_{1.4} L_{\rm{Edd}}
\approx
1.8\times10^{36}
m_{1.4}
\dot{m}_{16}
\,\,{\rm erg\;s^{-1}}
\label{eq:lx}
\end{eqnarray}
where $\epsilon_\mathrm{X}=GM_\mathrm{X}/(R_\mathrm{X} c^2) \approx 0.2$ 
represents the mass-to-energy conversion efficiency of the neutron star.
By comparing equation~(\ref{eq:ladaf}) with equation~(\ref{eq:lx}),
we note that $L_{\rm{X}}\gg L_{\rm{ADAF}}$, unless $\alpha$ is extremely small 
($\alpha\lesssim10^{-2}$).
This suggests that virtually all the observed X-ray luminosity arises from 
the polar cap regions of the neutron star.

In this section, as a model for normal X-ray outbursts, we have numerically 
studied the mass-accretion rate via RIAFs of material captured from 
truncated Be disks, for which there is a gap between the disk and 
the neutron star orbit. No analytical approach was adopted here, because
mass comes from a tidally elongated part of the disk, where the density and 
velocity distribution can be obtained only numerically. 
In the next section, where we consider a mechanism 
for giant X-ray outbursts, we will see that direct collision of the neutron star
with a Be disk extending beyond the neutron star orbit 
can also cause smaller-scale outbursts with the luminosity 
at a level of normal outbursts. In studying such interaction, we will adopt
an analytical approach as well as a numerical one for better understanding of 
physics at work there.

%
%
\section{Accretion Mechanism for Giant X-ray Outbursts}
\label{sec:type2}
%
In the previous section, we proposed that normal X-ray outbursts are 
caused by ADAFs of material transferred from a truncated Be disk. 
The peak mass-transfer rate obtained by numerical simulations
was in the range $10^{16-17}\,\mathrm{g\;s}^{-1}$.
This rate, together with the accretion timescale of RIAFs, gives rise to 
the peak X-ray luminosity of the order of $10^{36}\,\mathrm{erg\;s}^{-1}$.
As summarized below, however, it is about one order of magnitude lower than 
that of giant X-ray outbursts. In other words, 
the mass-transfer rate comparable to $10^{17-18}\,\mathrm{g\;s}^{-1}$ 
is required to explain the peak X-ray luminosity of giant X-ray outbursts.

At the first glance, it seems very unlikely for a truncated Be disk to provide 
such a high mass-transfer rate, because there is a gap between the 
neutron star orbit and the Be disk. 
However, in misaligned systems, where the tidal torque 
is weaker than in coplanar systems, the truncation radius 
could be larger than the periastron separation, depending on the tilt angle 
and the azimuth of tilt.
If such a disk is warped towards the orbital plane and if its direction happens 
to be near the periastron direction, the neutron star could capture gas 
at a high enough rate for a giant X-ray outburst, when it passages 
through the warped part of the Be disk. Indeed, disk warping episodes 
associated with giant X-ray outbursts have been observed,
as mentioned in section~\ref{sec:intro} and 
also summarized in section~\ref{subsec:type-ii}.

There are two possible mechanisms to make a Be disk warped; 
the radiation driven warping \citep{Pringle1996,Porter1999} and the tidally 
driven warping \citep{Martin2011}. Unfortunately, our SPH code 
cannot currently handle either of these mechanisms,
although there are SPH codes in which one of these mechanisms is implemented 
(\cite{Foulkes2010} for radiation-driven warping simulations; 
\cite{Lodato2010} for tidal warping simulations).
Hence, as a first step towards modeling giant X-ray outbursts, 
we study below the interaction between the neutron star and 
a rigidly tilted disk that extends beyond the neutron star orbit. 
After summarizing observational characteristics of giant X-ray outbursts, 
we first consider the Bondi-Hoyle-Lyttleton accretion of Be-disk material 
as a simplest approach
for modeling the passage of the neutron star through the Be disk, 
and then numerically study accretion flows from 
differentially rotating Be disks for more detailed modeling. 
Whereas numerical simulations enable us to quantitatively compare results,
the analytical approach provides us qualitatively better understanding of physics behind
the numerical result.
We also discuss the possibility that the accretion flows become 
supercritical if the Be disk density is very high.

%
\subsection{Observational Characteristics of Giant Outbursts}
\label{subsec:type-ii}
%
The observed features of giant X-ray outbursts are summarized as follows:
The X-ray luminosity increases by a factor $\gtrsim 10^2$ with respect to 
the quiescent state ($L_X \gtrsim 10^{37}\;\mathrm{erg\;s^{-1}}$). 
It sometimes rises up by a factor of $10^{3-4}$ to a level comparable with
the Eddington luminosity of the neutron star. 
The duration time is several tens of days 
($\gtrsim 0.5 P_{\mathrm{orb}}$, sometimes over one orbital period).
No orbital modulation has been detected.
The fact that the large spin-up of the neutron star is
seen in giant X-ray outbursts provides a strong evidence for the presence
of a transient accretion disk during the giant outburst.
Quasi-periodic oscillations (QPOs) detected in some systems 
(e.g., \cite{Takeshima1994,Finger1996,Heindl1999}) 
also support the transient accretion disk scenario.

In some systems, the warping of Be disk associated 
with the giant outburst has been observed 
\citep{Negueruela2001b,Reig2007,Moritani2011}.
\citet{Negueruela2001b} and \citet{Reig2007} has made long-term monitoring observations of
the H$\alpha$ emission line profile from the optical counterpart of 
4U~0115$+$634 in 1995--2005, and found that before and during giant X-ray outbursts, 
it changed from usual double-peaked profile to a single-peaked or
shell-line profile on a timescale of a year or so.
As mentioned in section~\ref{sec:intro}, this profile variability is most likely 
to originate from the precession of a warped Be disk.
Moreover, high-resolution H$\alpha$ profiles of the optical counterpart of A~0535$+$262 
observed during the giant X-ray outburst in 2009 had an enhanced component, which together with 
the double-peaked component, 
appeared as a triple peak.
Further analysis of these profiles indicates that the Be disk became tidally warped \citep{Martin2011} 
and intersected with the binary orbital plane near the periastron during the giant X-ray outburst (Moritani et al. 2012, in prep).

%
\subsection{Bondi-Hoyle-Lyttleton Accretion}
\label{subsec:bhl}
%
We consider the Bondi-Hoyle-Lyttleton (BHL) accretion process 
where the neutron star captures gas from an initially uniform 
flow of density $\rho$ and relative speed $v_\mathrm{rel}$. 
The BHL accretion was first analyzed by \citet{bh44} and have been 
extensively studied since then 
(see \cite{Edgar2004} for references and a modern review).

In the classical BHL accretion, the material with initial impact parameter smaller than a critical value 
\begin{equation}
b \equiv \frac{2GM_X}{v_\mathrm{rel}^{2}} 
  \approx 3.7 \times 10^{12}\,m_{1.4}
  \left( \frac{v_\mathrm{rel}}{100\,\mathrm{km\;s}^{-1}}
  \right)^{-2}\,\mathrm{cm}
\label{eq:bhl-radius}
\end{equation}
has negative total energy, and is eventually accreted on to the central object.
This leads to a simple BHL formula for the expected mass accretion rate,
\begin{eqnarray}
{\dot M}_0 &=& \rho v_\mathrm{rel} \pi b^{2} 
   \nonumber\\
   &=& 4.4 \times 10^{18} m_{1.4}^2
   \left( \frac{\rho}{10^{-14}\,\mathrm{g\;cm}^{-3}} \right) \nonumber\\
   && \times \left( \frac{v_\mathrm{rel}}{100\,\mathrm{km\;s}^{-1}}
   \right)^{-3}\,\mathrm{g\;s}^{-1}
\label{eq:mdotbhl}
\end{eqnarray}

\begin{figure}[!t]
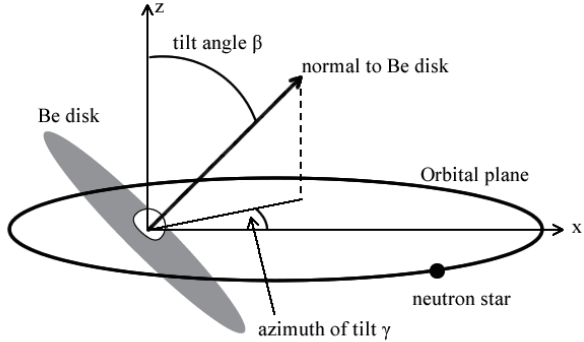

  \centering
  \FigureFile(0.45\textwidth,0.45\textwidth){figure6.eps} 
 \caption{
Schematic diagram of a misaligned Be/X-ray binary.
}
 \label{figure6}
\end{figure}

In order to emulate the accretion process from an extended tilted disk, we adopt the disk gas
density and velocity evaluated at the position of the neutron star, assuming the Be disk 
to cross the orbit at this position
(see figure~\ref{figure6} for the model geometry).
In the calculation, the Be disk is assumed to rotate at the Keplerian speed and 
have the density distribution in the form,
\begin{equation}
\rho = \rho_0 \times \left(\frac{r}{R_*}\right)^{-n}
\label{eq:rho_disk}
\end{equation}
with typical Be disk parameters, $\rho_0 = 10^{-11} \mathrm{g\;cm^{-3}}$ and $n = 7/2$.
Here, $r$ is the distance from the Be star.
For simplicity, we neglect the velocity shear and density gradient around the neutron star.

%

%
\begin{figure*}[!ht]
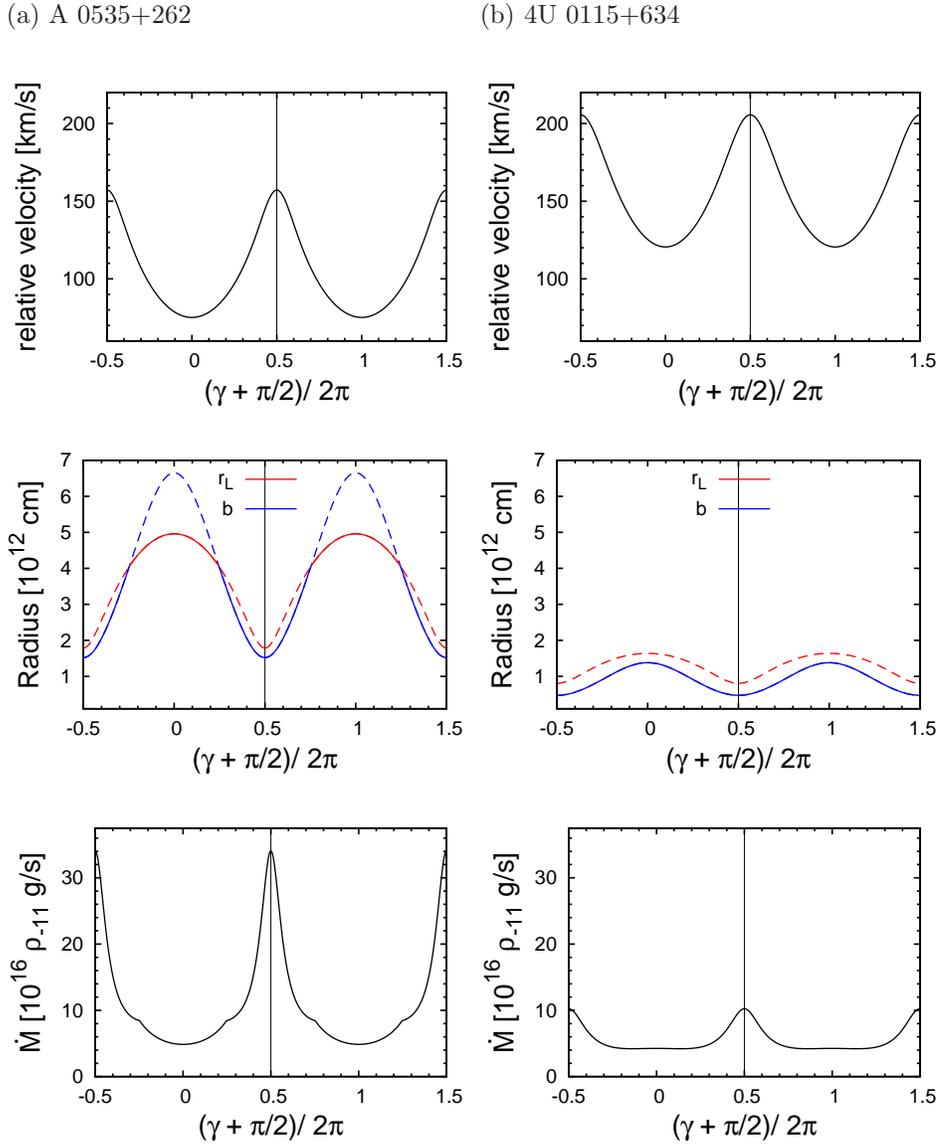

  \centering
  \begin{tabular}{p{0.35\textwidth}@{}p{0.35\textwidth}}
  (a) A~0535$+$262 & (b) 4U~0115$+$634 \\
  \FigureFile(0.35\textwidth,0.35\textwidth){figure7a-1.eps} &
  \FigureFile(0.35\textwidth,0.35\textwidth){figure7b-1.eps} \\
  \FigureFile(0.35\textwidth,0.35\textwidth){figure7a-2.eps} &
  \FigureFile(0.35\textwidth,0.35\textwidth){figure7b-2.eps} \\
  \FigureFile(0.35\textwidth,0.35\textwidth){figure7a-3.eps} &
  \FigureFile(0.35\textwidth,0.35\textwidth){figure7b-3.eps} \\
  \end{tabular}
 \caption{
 Characteristic quantities of the BHL accretion
 as a function of ascending-node azimuth $\gamma+\pi/2$,
 at which the neutron star passes through the Be disk: 
 (a) A~0535$+$262 and (b) 4U~0115$+$634.
 {\it Top panels}: Relative velocity between the neutron star and the Be disk.
 {\it Middle panels}: Comparison between BHL accretion radius $b$ (dotted line) and 
 Roche radius $r_{\mathrm{L}}$ (dashed line) of the neutron star.
 {\it Bottom panels}: Mass-accretion rate $\dot{M}_{\mathrm{BHL}}$ 
 given by equation~(\ref{eq:mdotbhl3}).
 In these calculations, the tilt angle $\beta$ is fixed at $45^\circ$.
 The base density and the density gradient index of the Be disk 
 are assumed to be  $10^{-11} \mathrm{g\;cm^{-3}}$ and 7/2, respectively.
 }
 \label{figure7}
\end{figure*}

As an example, in the top panels of figure~\ref{figure7}, 
we show the relative speed $v_{\mathrm{rel}}$ 
between the neutron star and a Be disk with tilt angle $\beta=45^{\circ}$
as a function of azimuth of the ascending node $\gamma + \pi/2$.
The left panel is for A~0535$+$262, while the right panel is for 4U~0115$+$634. In both calculations,
we assumed that the disk rotates in the prograde direction.
Note that the plot shows the dependence of the relative velocity on the azimuth of tilt, $\gamma$,
of the Be disk, not on the orbital phase of the neutron star. 
For $\gamma+\pi/2=0$, the neutron star passes through the Be disk
at the apastron, while for $\gamma+\pi/2=\pi$ the periastron is the location of the disk transit.

We mentioned above that the material with initial impact parameter smaller than $b$ accretes to the neutron star.
In binaries, however, only material inside the Roche radius of the neutron star, $r_{\rm L}$, given by equation~(\ref{eq:roche}), 
can be accreted. 
Thus, in the BHL accretion in binaries, the accretion rate is given by
\begin{equation}
\dot{M}_{\mathrm{BHL}} = \dot{M}_0 \times 
        min\left[ 1,\; \left( \frac{r_\mathrm{L}}{b} \right)^2 \right],
\label{eq:mdotbhl3}
\end{equation}
where $\dot{M}_0$ is the classical BHL accretion rate given by equation~(\ref{eq:mdotbhl}).
The middle panels of figure~\ref{figure7} compare these two radii, $b$ (blue line) and 
$r_{\mathrm{L}}$ (red line).
The solid part of these lines indicates the effective accretion radius, i.e., the smaller radius of $b$ and $r_{\mathrm{L}}$, 
inside which the gas accretes on to the neutron star.
We note that for $\beta = 45^\circ$, the accretion radius in A~0535$+$262 is controlled by $b$
when the neutron star passes through the disk around the periastron, while it is equal to $r_\mathrm{L}$ otherwise.
In our model, this transition occurs only for moderate values of tilt angle. 
This is also the case in 4U~0115$+$634, 
although in this smaller system with higher relative velocity, the accretion radius 
in the particular case of $\beta=45^\circ$ is controlled by $b$ for any value of azimuth of tilt $\gamma$.
For $\beta \ll 45^\circ$, because of small relative velocity, the critical radius $b$ for the BHL accretion 
exceeds the Roche radius $r_\mathrm{L}$ at any azimuth of the disk plane,
so that the accretion radius is always given by $r_\mathrm{L}$. On the other hand, 
for $\beta \gg 45^\circ$ (including all cases of retrograde disk rotation), the relative velocity is so high that
the BHL accretion radius $b$ is always smaller than the Roche radius $r_\mathrm{L}$,
as in the case of stellar wind accretion.

The bottom panels of figure~\ref{figure7} present
the accretion rate $\dot{M}_{\mathrm{BHL}}$ calculated by using equation (\ref{eq:mdotbhl3}).
Note that in A~0535$+$262, the mass-accretion rate of $\gtrsim 10^{17}\;\mathrm{g\;s}^{-1}$, 
and hence the accretion luminosity of 
$\gtrsim 10^{37}\;\mathrm{erg\;s}^{-1}$, is obtained for a wide range of
azimuth of the misaligned disk.
In contrast, in 4U~0115$+$634, such a high accretion rate is obtained only if 
the Be disk density is significantly higher than the typical density.

%
\begin{figure}[!t]
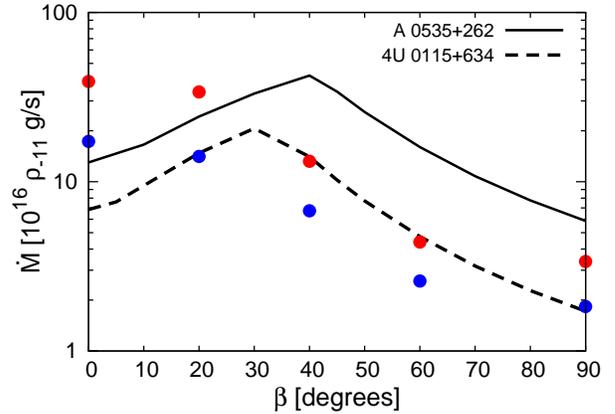

\centering
\FigureFile(0.45\textwidth,0.45\textwidth){figure8.eps}
\caption{
 Mass-accretion rate as a function of tilt angle $\beta$.
 The solid and the dashed lines denote analytical $\dot{M}_\mathrm{BHL}$ 
 for A~0535$+$262 and 4U~0115$+$634, respectively, while the red and the blue
 circles denote the peak mass-accretion rate 
 obtained by numerical simulations of A~0535$+$262 and 4U~0115$+$634, respectively.
 In these calculations, the azimuth of tilt, $\gamma$, is fixed at $90^\circ$, for which
 the neutron star passes through the Be disk at the periastron.
}
\label{figure8}
\end{figure}

Figure~\ref{figure8} shows the mass accretion rate as a function of tilt angle $\beta$.
In these calculations, the azimuth of tilt $\gamma$ was fixed at $90^\circ$, for which
the neutron star passes through the Be disk at the periastron. 
The solid and the dashed lines denote $\dot{M}_\mathrm{BHL}$ for A~0535$+$262 and 4U~0115$+$634, respectively.
From the figure, we note that in A~0535$+$262 
$\dot{M}_\mathrm{BHL} \gtrsim 10^{37}\;\mathrm{erg\;s}^{-1}$ is obtained 
for a very wide range of tilt angle for typical Be disk parameters ($\rho_0=10^{-11}\,\mathrm{g\;cm}^{-3}$ 
and $n = 7/2$), 
while in 4U~0115$+$634 it is reached only for a small range of $\beta$ as long as 
the typical Be disk parameters are adopted.
This difference arises from the fact that
larger systems (A~0535$+$262 in our sample systems) have 
larger Roche radii, which make the accretion rate higher. 
From figure~\ref{figure8}, we also note that
the highest accretion rate occurs at $\beta \sim 30-40^{\circ}$.
This angle is slightly smaller in 4U~0115$+$634 than in A~0535$+$262.
This is because
small systems have relatively larger relative velocity.

In this subsection, we have considered the BHL accretion
as a model for the direct accretion from the misaligned Be disk.
However, the Be disk is differentially rotating, so that accretion occurs in a sheared flow.
The finite thickness of the disk and the eccentric orbit of the neutron star 
should also affect the accretion rate.
Therefore, the BHL accretion assuming a uniform background flow 
should be taken as a first order model. 
In the next section, we numerically examine accretion flows 
in differentially rotating Be disks, using SPH simulations.

%
\subsection{Accretion from Differentially Rotating Be Disks}
\label{sec:afsf}
%

Although very simplified, the above analytical model provides a useful information
about the mass captured by the neutron star when it passes through a tilted Be disk.
The resultant mass-capture rate is of the order of $10^{17}\rm{erg\;s^{-1}}$ 
for a wide range of tilt angles. 
This rate is comparable to the mass accretion rate required for 
the observed luminosity of giant X-ray outbursts.

As seen in section~\ref{sec:sph-results-a0535}, however, high mass-capture rates 
do not necessarily mean high mass-accretion rates.
If the accretion time is comparable with or longer than the orbital period, the accretion luminosity
will stay more or less constant at a level much lower than that calculated from 
the peak mass-capture rate.
Hence, it is important to examine the accretion timescale of mass captured by the neutron star.
It is, however, not a trivial task to estimate the accretion rate and timescale, 
given the presence of differential rotation in the Be disk and 
the eccentricity of the neutron star orbit.
For quantitative study of these quantities, 
we carried out SPH simulations of accretion processes from 
misaligned Be disks for a limited range of parameter space for A~0535$+$262 
and 4U~0115$+$634.

\subsubsection{Numerical Model}
\label{sec:sph-model2}

The initial configuration of simulations are as follows.
For simplicity, we consider a rigidly-tilted, isothermal Be disk to model the interaction
between the neutron star and a 
misaligned disk.
In order to simulate direct collision of the neutron star with the Be disk 
with good spatial resolution, we concentrated SPH particles in a ring-like region
across the periastron separation $r = (1-e)a$,
from $r \sim 0.65 (1-e)a$ to $r \sim 1.2 (1-e)a$.
We distributed SPH particles such that the Be disk has the equatorial density distribution 
in the form of equation~(\ref{eq:rho_disk}) with $\rho_0=10^{-11}\,\mathrm{g\;cm}^{-3}$ 
and $n=7/2$ and is in hydrostatic equilibrium in the direction normal to 
the disk mid-plane.
The disk temperature is set to $0.6\,T_\mathrm{eff}$ as in section~\ref{sec:sph-code}.
We tilted the Be disk about the semi-major axis (i.e., $\gamma=90^\circ$)
to study the strongest interaction case for each tilt angle.
The accretion radius of the neutron star is set to 
$0.01a$ $\approx 2 \times 10^{5} R_\mathrm{X}$.
The initial number of SPH particles is $7 \times 10^5$.
We have run each simulation from $0.1 P_\mathrm{orb}$
prior to the periastron through $0.1 P_\mathrm{orb}$ after it.

\subsubsection{Numerical Results}
\label{sec:sph-result2}

We have run simulations for five different tilt angles 
$\beta= 0^\circ$, 20$^\circ$, 40$^\circ$, 60$^\circ$, and 90$^\circ$
for each system.
The peak accretion rates obtained in these simulations are plotted by filled circles in 
Figure~\ref{figure8}.
In contrast to the analytical result, the simulated peak accretion rate 
monotonically decreases with increasing misaligned angle. 
Compared with the analytical rate, the simulated rate is significantly high
in the coplanar ($\beta=0$) case, comparable for $\beta=20^\circ$ and $90^\circ$, 
and significantly low for $\beta=40^\circ$ and $60^\circ$.

\begin{figure}[!t]
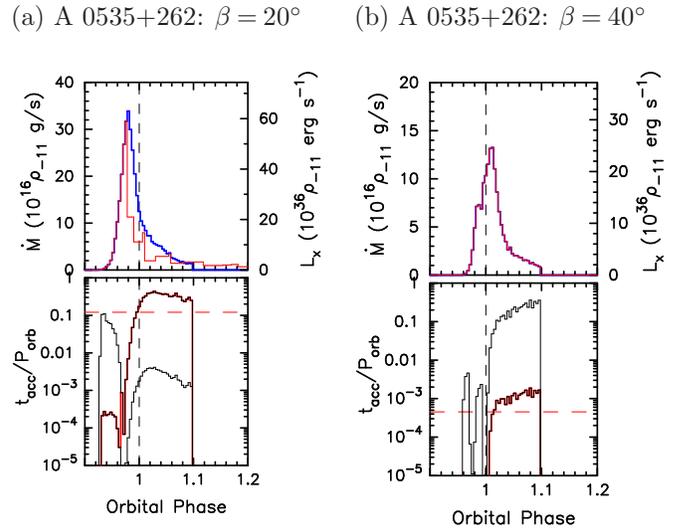

\centerline{
\begin{tabular}{p{0.23\textwidth}p{0.23\textwidth}}
(a) A~0535$+$262: $\beta=20^\circ$ & (b) A~0535$+$262: $\beta=40^\circ$ \\
\FigureFile(0.23\textwidth,0.23\textheight){figure9a.eps} &
\FigureFile(0.23\textwidth,0.23\textheight){figure9b.eps} 
\\
\end{tabular}}
 \caption{Simulation result for the BHL accretion from a tilted Be disk in A~0535$+$262:
 (a) $\beta=20^\circ$ and (b) $\beta=40^\circ$, where $\beta$ is the tilt angle.
 The azimuth of tilt $\gamma$ is fixed at $90^\circ$, for which
 the neutron star passes through the Be disk at the periastron.
 Each simulation has run from orbital phase $-0.1$ to $+0.1$.
 {\it Upper panels}: The orbital-phase dependence of the mass-inflow rate 
 through the inner simulation boundary
 [thick (blue) lines] and the estimated accretion rate at 
 the neutron star radius [thin (red) lines)].
 The right axis shows the X-ray luminosity corresponding to these rates.
 {\it Lower panels}: The orbital-phase dependence of 
 the accretion timescale [thick (red) lines].
 The standard and the ADAF accretion timescales are also shown by the upper and 
 the lower thin (black) lines, respectively.
 The horizontal (red) dashed line in each lower panel denotes 
 the mass-weighted average of the accretion timescale.
 }
 \label{figure9}
\end{figure}

\begin{figure}[!ht]
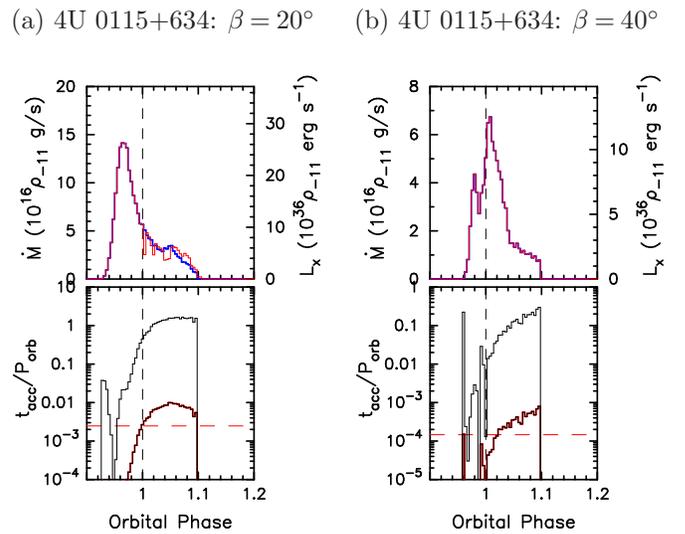

\centerline{
\begin{tabular}{p{0.23\textwidth}p{0.23\textwidth}}
(a) 4U~0115$+$634: $\beta=20^\circ$ & (b) 4U~0115$+$634: $\beta=40^\circ$ \\
\FigureFile(0.23\textwidth,0.23\textheight){figure10a.eps} &
\FigureFile(0.23\textwidth,0.23\textheight){figure10b.eps} 
\\
\end{tabular}}
 \caption{Simulation result for the BHL accretion from a tilted Be disk in 4U~0115$+$634:
 (a) $\beta=20^\circ$ and (b) $\beta=40^\circ$.
 The azimuth of tilt $\gamma$ is fixed at $90^\circ$, for which
 the neutron star passes through the Be disk at the periastron.
 Each simulation has run from orbital phase $-0.1$ to $+0.1$.
 The format of the figure is the same as that of figure~\ref{figure9}.
 }
 \label{figure10}
\end{figure}

In order to understand the cause(s) of these differences, 
we compare simulation results
for tilt angles $\beta = 20^\circ$ and $\beta=40^\circ$ in more detail
in figures~\ref{figure9} (A~0535$+$262) 
and \ref{figure10} (4U~0115$+$634).
In each figure, upper panels show the orbital-phase dependence of 
the mass accretion rate on to the neutron star.
In each panel, the thick (blue) line denotes the rate of mass inflow 
through the inner simulation boundary (hereafter, the simulated accretion rate).

For $\beta=20^\circ$, the simulated accretion rate rises rapidly, has a pre-periastron peak, and 
then decays slowly. On the other hand, for $\beta=40^\circ$, there are two peaks: 
The lower pre-periastron peak is followed by the higher post-periastron peak. 
After the post-periastron peak, the rate decreases gradually.
In both cases, the initial accretion is of the BHL type, but later (after
the neutron star passes through the disk) the accretion is via an accretion disk 
formed by material that catches up with the neutron star from behind, 
which has significantly larger specific angular momentum than 
the martial accreted before and during the disk transit. 
The initial BHL accretion and the subsequent accretion via the accretion disk 
form the rapidly rising part and the slowly decline part of 
the simulated accretion rate profile, respectively.

As seen in figure~\ref{figure8}, 
the analytical model predicts a higher accretion rate for $\beta=30^\circ-40^\circ$ 
than for $\beta=20^\circ$. The simulations, however, 
have revealed the opposite trend. The simulated accretion rate 
for $\beta=20^\circ$ is 2--3 times as high as that for $\beta=40^\circ$
in both systems. 
Examining the interaction more closely, we have found that in the case of 
small tilt angles, the tidal torque by the neutron star starts exciting a tidal stream 
in the Be disk significantly before the disk crossing event, 
which collides with the neutron star at a very small impact parameter. Since the 
density in the tidal stream is much higher than in the background disk gas, 
this collision enhances the mass-accretion rate significantly. In contrast, 
for $\beta \gtrsim 40^\circ$, where the tidal torque is weaker, the tidal 
stream grows more slowly. By the time it becomes strong, the neutron star 
is already passing the periastron and the streaming gas, which has a 
relatively large impact parameter, forms an accretion disk, not directly 
running on to the neutron star, thus unable to significantly enhance 
the mass-accretion rate.

Since the accretion radius set in these simulations is 
$0.01 a \sim 2 \times 10^{5} R_\mathrm{X}$ and still way far from the neutron star,
the accretion time the flow takes from this radius to the neutron star surface
could significantly modify the amplitude and shape of the accretion rate profile.
Thus, we estimated the accretion rate at the neutron star radius 
(hereafter, the final accretion rate) using the same procedure described 
in section~\ref{sec:sph-results-a0535}.
In deriving the final accretion rate, we assumed that 
the accretion flow is of standard-disk type 
in the period of $\dot{m}>1$ (or $\dot{m}_{16}>20$) and afterwards,
while it is of ADAF type otherwise.
In the former (latter) case,
the accretion takes place in the viscous (ADAF) timescale given by equation~(\ref{eq:tvis2})
[equation~(\ref{eq:t_adaf})], 
with the circularization radius derived from the specific angular momentum of accreted SPH particles.
The phase dependence of the accretion timescale, $t_\mathrm{acc}$, 
is shown by the thick (red) lines in the lower panels of figures~\ref{figure9} and \ref{figure10}, where for comparison purpose, 
the viscous and ADAF timescales are also denoted by the upper and lower thin lines, 
respectively.
The resultant, final accretion rate is shown by the thin (red) lines 
in the upper panels.
From the figures we note that the final accretion rate profile 
changes little from 
the simulated profile at $r=0.01a$, except for cases where the accretion rate exceeds 
the upper limit for ADAF solutions, so that the accretion flow is of standard-disk type.
For typical Be disk parameters, the latter case
occurs only in $\beta \le 20^\circ$ simulations for A~0535+262.

The above simulations of accretion from differentially rotating Be disks 
have important observational implications. The peak accretion rates obtained 
for $\beta \lesssim 20^\circ$ and typical Be disk density ($\rho_{-11} \sim 1$) 
agree with those required for
the observed X-ray luminosity of giant X-ray outbursts. 
In other words, Be disks of typical density  
have potential to cause giant outbursts if the tilt angle is small. 
On the other hand, if the tilt angle is large, 
say $\beta \sim 40^\circ$, the Be disk has to be several times denser than typical  
in order to supply large enough amount of gas for giant X-ray outbursts.
Otherwise, the outbursts will appear as normal X-ray outbursts.

%
\subsection{Supercritical Accretion}
\label{subsec:scaf}
%
In the previous section, we assumed that the accretion 
of mass captured by the BHL-type accretion occurs via the standard disk or optically thin ADAF.
This assumption is reasonable as long as the mass-capture rate
is of the order of $10^{17}\,\mathrm{g\;s}^{-1}$ or less.
If the mass-capture rate exceeds the Eddington rate ($\sim 10^{18}\,\mathrm{g\;s}^{-1}$), however, 
the accretion flow drastically changes to being supercritical.
In this section, we investigate the possibility that Be/X-ray binaries have 
supercritical accretion flows. 

Recent X-ray observations have revealed the existence of bright X-ray sources
such as GRS~1915$+$105 in our galaxy with luminosities 
over the Eddington luminosity (c.f., \cite{done07}). 
Such large luminosities can be explained by a supercritical 
accretion flow on to a black hole or neutron star (c.f., \cite{wftm00}).

In an optically thick accretion flow with an accretion rate much higher
than the Eddington accretion rate, the photons are trapped and restored as
the entropy in the accreting gas without being radiated away.
Such a flow is called a slim disk, where 
the advective cooling dominates the radiative cooling.
The slim disk model was first proposed by \citet{Abramowicz88} as
a thermally stable, advection-dominated, supercritical accretion flow.
Then, its self-similar solution has been derived by \citet{wz99} and \citet{wa06}.

In what follows, we first describe the condition that the supercritical accretion occurs
in the accretion flow around the neutron star
in the context of Be/X-ray binaries.
We then derive the 
accretion timescale based on the self-similar solutions of slim disk model \citep{wa06}.
We finally evaluate the Bolometric X-ray luminosity emitted from the slim disk.

%
\subsubsection{Condition for Supercritical Accretion}
\label{sec:photon-trap}
%
At luminosities close to or higher than the Eddington luminosity,
the scale-height of the accretion disk is comparable to the radius, i.e., $H/r\sim1$.
Hence, in the following analysis, we do not distinguish between
spherically symmetric accretion and disk accretion.

In a very optically thick medium, photons are trapped 
if their diffusion timescale $t_\mathrm{diff}$ is
longer than the accretion timescale $t_\mathrm{acc}$. 
If the medium is disk-like, $t_\mathrm{diff}$ of photons near the equatorial plane
is given by
\begin{equation}
t_\mathrm{diff} \sim \frac{3H^2}{c \lambda} \sim \frac{3H\tau}{c} 
   \sim \frac{3 H \kappa \Sigma}{2c}
\label{eq:t_diff}
\end{equation}
(e.g., \cite{Kato2008}), where $\lambda$ is the mean free path of photons, 
$\tau$ is the vertical optical depth,
$\kappa$ is the opacity, and $\Sigma$ is the surface density.
In deriving the last equation, we used $\tau \sim \kappa \Sigma/2$.
As the opacity $\kappa$, we use that of the electron scattering, 
$\kappa_\mathrm{es}$ ($\sim 0.4$).

On the other hand, the accretion timescale is written as
\begin{equation}
t_\mathrm{acc} \sim \frac{r}{v_r} \sim 
\frac{2 \pi r^2 \Sigma }{\dot{M}}.
\label{eq:t_acc}
\end{equation}
Equating equation~(\ref{eq:t_diff}) with equation~(\ref{eq:t_acc}) and using
equation~(\ref{eq:L_Edd}), we have the photon-trapping radius $r_\mathrm{trap}$ as
\begin{equation}
r_\mathrm{trap}
=
\frac{3}{2}\frac{H}{r}
\frac{\dot{M}c^2}{L_{\rm{Edd}}}
r_{\rm S}
\approx
3.2\times10^{4}
\dot{m}_{16}
\,\,{\rm cm},
\label{eq:r_trap1}
\end{equation}
where $H/r \sim 1$ at high luminosities.
In order for the supercritical accretion to occur, 
$r_\mathrm{trap}$ should be larger than the magnetospheric radius $r_\mathrm{M}$, 
from which we have the following condition:
\begin{eqnarray}
\dot{m}_{16}
>
\dot{m}_{16, \rm sc},
\label{eq:mdot_trap}
\end{eqnarray}
where we define 
\begin{equation}
\dot{m}_{16, \rm sc}
=1.7\times10^{3}k^{7/9}m_{1.4}^{-1/9}\mu_{30}^{4/9}.
\label{eq:scc}
\end{equation}
From equation~(\ref{eq:scc}), the mass-capture rate must be higher 
than $\sim2\times10^{19}\rm{g\;s}^{-1}$ in order that the accretion flow becomes supercritical
before it reaches the magnetosphere of the neutron star.
To realize such a huge mass-capture rate, the base density of the Be disk should be
by two orders of magnitude higher than the typical $\rho_0=10^{-11}\rm{g\;cm}^{-3}$.
Note that this is not an unrealistic condition.
For instance, figure~\ref{figure9}(a) shows that in the case of A~0535+262 
with $\beta=20^{\circ}$ and $\gamma=90^\circ$, 
the accretion flow can be supercritical at orbital phases from $-0.04$ to 0.02 
with respect to the periastron, if the base density is $10^{-9}\,\mathrm{g\;cm}^{-3}$ 
or $\rho_{-11}=100$.

%
\subsubsection{Observational Implications}
%
In this subsection, we evaluate the Bolometric X-ray luminosity
emitted from the slim disk. We assume that the condition $\dot{m}>\dot{m}_{\rm sc}$
is satisfied so that the slim disk state holds from the photon-trapping
radius down to the magneto-spheric radius.

 According to a self-similar solution of the slim disk \citep{wa06},
 the radial velocity is given by
 \begin{eqnarray}
 |v_{r}|\approx3.2\times10^9
 f\alpha_{0.1}\hat{r}^{-1/2}
 \,{\rm cm\;s}^{-1},
 \label{eq:vr_sc}
 \end{eqnarray}
 where we recall that $f=Q_\mathrm{adv}/Q^{+}$.
 Then, the accretion timescale is calculated as
 \begin{equation}
 t_{\rm{SCAF}}=\frac{r_{\rm{trap}}}{|v_r|}
 \approx
 2.7\times10^{-6}
 f^{-1}\alpha_{0.1}^{-1}
 m_{1.4}^{-1/2}
 \dot{m}_{16}^{3/2}
  \,\,\,{\rm s}.
 \label{eq:acctime_sc}
 \end{equation}
The distribution of effective temperature of the slim disk 
is given by the self-similar solution as
\begin{eqnarray}
T_{\rm{eff}}
\approx 
4.9\times10^7
f^{1/8}
m_{1.4}^{-1/4}
\hat{r}^{-1/2}
\,\,\rm{K}
\end{eqnarray}
\citep{wa06}, so that
the cooling rate of the slim disk $Q_{\rm rad}$ is given by
\begin{eqnarray}
Q_{\rm{rad}}
&=&
2\sigma T_{\rm{eff}}^4
\nonumber \\
&\approx&
1.1\times10^{26}
f^{1/2}
m_{1.4}^{-1}
\hat{r}^{-2}
\,\,{\rm erg\,s^{-1}\,cm^{-2}}.
\end{eqnarray}
The Bolometric X-ray luminosity emitted from the slim disk is then estimated as
\begin{eqnarray}
L_{\rm{slim}}
&=&
2\pi{r_{\rm S}^2}\int_{\hat{r}_{\rm M}}^{\hat{r}_{\rm trap}}\hat{r}Q_{\rm{rad}}d\hat{r}
\nonumber \\
&\approx&
1.2\times10^{38}
f^{1/2}
m_{1.4}
\ln\left(\frac{\hat{r}_{\rm trap}}{\hat{r}_{\rm M}}\right)
\,\,{\rm erg\,s^{-1}}.
\label{eq:lslim}
\end{eqnarray} 

From equation~(\ref{eq:acctime_sc}), the timescale of supercritical accretion is much shorter than 
the orbital period of any Be/X-ray binaries. 
Hence, if the mass-capture rate satisfies the condition 
given by equation~(\ref{eq:mdot_trap}), 
the resultant supercritical accretion flow will have
the same final accretion rate profile as those 
shown by blue lines in the upper panels of 
figures~\ref{figure9} and \ref{figure10} 
with the amplitude proportional to the base density of the Be disk.
From equation~(\ref{eq:lslim}), the luminosity is then estimated to 
be of the order of $10^{38}\rm{erg\;s^{-1}}$ even if $\hat{r}_{\rm trap}/\hat{r}_{\rm M}\sim10$.  
Note that the light curves of supercritical accretion flows in Be/X-ray binaries 
are thus quite different from those expected for standard disk accretion.

%
\section{Discussion} 
\label{sec:discussion}
%
In Be/X-ray binaries, the Be disk is tidally/resonantly truncated by 
the gravitational force of the neutron star, except in highly eccentric 
systems and those with Be disks highly inclined from the orbital plane 
\citep{Okazaki2001}. Such a truncation produces a gap between the 
Be disk and the neutron star orbit. 
In a system with a gap, the neutron star captures gas from a tidally 
elongated part of the disk. Since the tidal deformation is a dynamical process 
and there occurs no redistribution of the angular momentum by viscosity, 
the rotation velocity in the elongated part decreases more rapidly with 
radius than in a Keplerian disk. This makes the relative velocity 
between the neutron star and the captured material larger than in the 
case where the neutron star captures mass from an extended Keplerian disk
without a gap.
As a result, the material captured from a truncated Be disk
forms a large accretion disk around the neutron star. 
If the accretion disk is radiative-cooling dominated, 
its accretion timescale easily exceeds the orbital period.
In such a situation the system will exhibit no rapid nor large 
X-ray flux changes seen in normal X-ray outbursts.
In fact, in section~\ref{sec:sadm}, we have shown that this is the 
case for the Be/X-ray binaries A~0535$+$262 and 4U~0115$+$634, by using results 
from numerical simulations.

One might argue that the magnetosphere of the neutron star controls accretion flows, 
causing rapid and large variation of X-ray luminosity. 
When the magnetospheric radius is larger than the coronation radius 
of the accreting matter, the matter is expelled by the rotating magnetosphere 
(the propeller effect) and the systems is in the quiescent state. 
The gate opens only if the Be disk supplies large enough mass 
so as to the resultant accretion flow can push back the magnetosphere 
inside the corotation radius. In fact, in 4U~0115$+$63, a sudden transition 
from the quiescence to a normal outburst has been witnessed, 
which is consistent with this idea \citep{Campana2001}. 

However, as shown in section~\ref{sec:sadm}, A~0535$+$262 
is almost always in the direct accretion regime (figure~\ref{figure3}), 
because the propeller effect does not work for this slowly rotating X-ray pulsar.
The system is therefore one of best candidates to study more direct 
relationship between the accretion flows on to the neutron star and 
the resultant X-ray variability. Given that standard disks have accretion 
timescales much longer than the orbital period, we can safely rule out the 
possibility that the material from the truncated Be disk forms the standard disk 
around the neutron star.

The problem with rapid and large X-ray variability is resolved if the accretion flow is
radiatively inefficient (RIAF).
In section~\ref{sec:type1}, using results from numerical simulations, we have shown that 
mass transfer from truncated Be disks with typical density parameters gives rise to accretion rates 
in agreement with those expected from X-ray luminosity of normal outbursts, and that
these rates are significantly lower than the critical rate above which no RIAF solution exists.
The accretion timescale is also consistent with the duration of normal outbursts.

The RIAF model for normal X-ray outbursts opens a new research field on Be/X-ray binaries.
In this paper, we assumed that the magnetospheric radius for self-similar ADAF solutions
is about the same as that for disk accretion. Given the magnetospheric interaction affects 
the accretion on to the neutron star in many ways, 
including even the possibility of launching jets, it is very important to investigate it
in detail. There is also a possibility to detect high energy emission from RIAFs,
e.g., synchrotron emission in X-rays from accelerated particles in the layer 
above the magnetosphere and the GeV/TeV emission via the inverse Compton scattering 
of soft photons from the Be star and/or the accretion disk, although \emph{Fermi} and VERITAS 
have detected no emission above 0.1\,GeV from A~0535$+$262 during the 2009 giant outburst 
\citep{Acciari2011}.
We will study it in a forthcoming paper.

Another interesting possibility also emerges from the RIAF model for normal X-ray outbursts. 
In section~\ref{sec:type1}, we have used the numerical results for a typical Be disk 
with base density of $10^{-11} \mathrm{g\;cm}^{-3}$ and density gradient index $n = 7/2$. 
There is, however, a large range of observed base density 
($10^{-12} \mathrm{g\;cm}^{-3} - 10^{-10} \mathrm{g\;cm}^{-3}$) and 
density gradient index (2.5--4.5) (e.g., \cite{Jones2008}). 
If the mass transfer rate from the Be disk exceeds 
the critical rate above which no RIAF solution exists, a standard accretion disk would form.
Thus, the transition between standard disk state and RIAF state can occur,
depending on the base density and density gradient of the Be disk.
Observationally, this transition will be seen as a sudden drop/rise of X-ray luminosity by 
about one order of magnitude. Since the mass-transfer rate decreases with increasing angle
between the Be disk and the binary orbital plane,
the coplanar Be-disk systems are most favorable to see this transition.

In section~\ref{sec:type2}, we have proposed that 
giant X-ray outbursts can be triggered by the BHL accretion of material 
from a tilted Be disk. Although the RIAF model with truncated Be disks works 
for normal X-ray outbursts, it cannot be applied to giant X-ray outbursts by two reasons. 
One reason is that because of the disk truncation, the mass-transfer rate
is lower than that required for giant X-ray outbursts, unless the Be disk density is
much higher or its distribution is much flatter than typical values. 
The other, more crucial reason is that the mass-accretion rate 
estimated from the luminosity of bright giant X-ray outbursts exceeds by a factor of few 
the highest mass-accretion rate for RIAFs. 
Moreover, there is observational evidence that the Be disk is warped when
giant outbursts occur \citep{Negueruela2001b, Reig2007, Moritani2011}.

Our model for giant X-ray outbursts is based on the idea that 
the warped shape of the Be disk naturally results in the enhanced 
mass accretion, if the 
disk is inclined from the orbital plane 
and its warped part gets across the orbit of the neutron star from 
above or below the orbital plane. 
In real systems, the accretion rate should depend on the local 
density and velocity distribution of gas around the neutron star. 
For a warped disk picture, however, no such information is easily available.
Therefore, as a first order of approximation, we have modeled 
the warped Be disk by a rigidly-tilted, Keplerian disk with a 
power-law density distribution in the radial direction.
The Be disk was assumed to have an outer radius larger than 
the periastron separation. With a simplified analytical treatment, 
we found that 
the accretion radius of the neutron star is limited by the Roche lobe radius for small 
tilt angles, while for large tilt angles the neutron star accretes material 
inside the BHL accretion radius. 

Numerical simulations also revealed a new feature 
of the BHL accretion in Be/X-ray binaries. For Be disks 
with small tilt angles ($\beta \lesssim 20^\circ$), the tidal 
torque of the neutron star rapidly excites a collimated gas flow 
in the Be disk, and the neutron star collides with this tidal stream 
at a very small impact parameter. Since the density in the tidal 
stream is much higher than in the background disk gas, this 
collision greatly enhances the mass-accretion rate. This does 
not happen for larger tilt angles ($\beta \gtrsim 40^\circ$). 
Consequently, the accretion rate 
is much higher for small tilt angles than for large tilt angles.

Using simulation data, we have estimated the time-dependent accretion rate 
at the neutron star radius. It turns out that for typical Be disk parameters, 
only long period systems, such as A~0535$+$262, with small 
tilt angles ($\beta \lesssim 20^\circ$) can have mass-accretion rates 
as high as those estimated for giant X-ray outbursts. In order for systems with 
large tilt angles ($\beta \gtrsim 40^\circ$) and/or short periods such as 4U~0115$+$634 
to exhibit giant X-ray outbursts, the Be disk density has to be higher than 
the typical one by a factor of several or more around the neutron star orbit.
Otherwise, the outburst luminosity 
of these systems falls into the range of normal X-ray outbursts.

In this paper, we have assumed the upper limit for the ADAF solutions to be 
$\dot{m} \equiv \dot{M}/(L_\mathrm{Edd}/c^2) = 1$ (or $\dot{M} \sim 2 \cdot 10^{17}\,\mathrm{g\;s}^{-1}$).
With this criterion, the accretion flow is radiative-cooling dominated (i.e., a standard disk)
only for relatively bright 
($\gtrsim 4 \cdot 10^{37}\,\mathrm{erg\;s}^{-1}$) giant outbursts.
For less bright ($\lesssim 4 \cdot 10^{37}\,\mathrm{erg\;s}^{-1}$) giant outbursts
and all normal outbursts, the accretion flow is advection-dominated.
The type of the accretion flow, standard-disk type or ADAF type, thus
depends only on whether the mass-accretion rate is higher than the limiting rate for ADAF
solutions or not.

From the sensitiveness of the accretion rate on the \emph{local} tilt angle, 
a working hypothesis comes up for a series of X-ray activities consisting of 
a few normal X-ray outbursts and a giant X-ray outburst. This series of 
X-ray activity has sometimes been observed in Be/X-ray binaries. As 
mentioned in section~\ref{subsec:type-ii}, 4U~0115$+$634 has shown 
observational evidence of precession of a warped Be disk before and 
during giant X-ray outburst \citep{Negueruela2001b, Reig2007}. 
When a warped disk precesses, it is likely that the local tilt angle, i.e., 
the angle between the neutron star's motion and the motion of material 
in the Be disk, varies from cycle to cycle. If the tilt angle is large at the 
beginning, the system exhibits an X-ray outburst at a level 
considered to be a normal outburst.
If the warped structure survives over several orbital cycles, there will be 
a good chance for the neutron star to pass through the disk at a small 
tilt angle in due course. This encounter will give rise to a giant X-ray outburst. 
Later on, the system will show a few outbursts at the level of normal X-ray 
outbursts until the azimuth of the disk plane ($\gamma+\pi/2$) moves
too far away from the periastron. In this way, our model for giant X-ray 
outbursts naturally, though qualitatively, explains a sequence of X-ray 
outbursts, i.e., the occurrence of a few normal X-ray outbursts before 
and after a giant X-ray outburst, which
has been observed in 4U~0115$+$634, A~0535$+$262, and some 
other Be/X-ray binaries.

Finally, there is possibility of supercritical accretion in Be/X-ray binaries,
as discussed in section~\ref{subsec:scaf}. Throughout the paper, we fixed the base density
and the density gradient index of the Be disk at $10^{-11}\,\mathrm{g\;cm}^{-3}$ and 7/2, 
respectively.
As mentioned previously, 
however, there is a large scatter in the distribution of the Be disk 
base density and its density gradient.
For Be disks with very high base density and/or significantly flatter density distribution, 
the model predicts that the accretion becomes supercritical, 
triggering super-Eddington luminosity. As long as we know, 
no Be/X-ray binaries have shown super-Eddington luminosity.
Nevertheless, we see no mechanism that can prevent Be disks 
from supplying mass at supercritical rates. If the supercritical 
accretion takes place, the luminosity will exceed or be comparable to 
the Eddington luminosity and photon trapping will veil all features related 
to magnetic fields of the neutron star such as X-ray pulsation and cyclotron 
resonance scattering features. Once such characteristic changes are observed, 
a fertile research field will arise from it.
It is therefore important to carry out more detailed modeling of 
supercritical accretion in Be/X-ray binaries. 
Be/X-ray binaries thus provide a unique opportunity where in one system
one can study physics of various types of accretion, 
including standard disks, Bondi-Hoyle-Lyttleton accretion, RIAFs, 
and even supercritical accretion flows, 
in relation to the geometry and dynamics of the mass donor.
%

%
\section{Conclusions}
\label{sec:conclusions} 

We have studied the origins of 
normal and giant X-ray outbursts in Be/X-ray binaries.
Based on the results from 3D SPH simulations,
we first showed that the accretion via the standard disk cannot
produce the observed rapid and large X-ray flux variation. 
Then, by using analytical studies as well as numerical ones, 
we proposed the following scenarios for these two types of X-ray outbursts:
\begin{enumerate}
\item The normal X-ray outbursts are caused by RIAFs from tidally truncated Be disks. 
The resultant luminosity and time variability are consistent with those observed, 
because RIAFs have accretion timescale much shorter than 
the orbital period, so that the rapid and large variation in the mass-transfer rate
from the truncated Be disk is conserved during the accretion process.
\item The giant X-ray outbursts occur in systems where 
the Be disk is misaligned with the binary orbital plane and 
is sufficiently developed. 
When the outermost part of such a disk is warped and crosses 
the orbit of the neutron star near the periastron,
the neutron star can capture a large amount of gas via the BHL accretion. 
This process results in the formation of a very small standard disk or ADAF,
depending on the mass-capture rate,
in which the accretion timescale is much shorter than the orbital period.
\end{enumerate}

In this paper, we focused on the mass-transfer process from the Be disk to the neutron star
and the subsequent accretion processes.
There are, however, observational hints that the giant X-ray outburst
is not a single independent event, but is a highlight of a much longer cycle of events.
For instance, a giant X-ray outburst is often accompanied by smaller scale X-ray outbursts,
and the Be disk emission starts declining sometime before the giant outburst.
In the next paper, we will study how the giant X-ray outburst and these associated events
are related to the evolutionary cycle of the Be disk.

\bigskip

This work has started, being inspired by discussions at the Be/X-ray Binary 2011 Workshop,
held in Valencia, 11-14 July 2011. The authors thank the organizers and participants of the workshop
for the exchange of stimulating ideas.
They also thank Kazuo Hiroi for his constructive comment on X-ray data.
The SPH simulations were performed on HITACHI SR16000 models
at the Information Initiative Center (iiC), Hokkaido University and at Yukawa Institute of
Theoretical Physics, Kyoto University. This work
was partially supported by the iiC collaborative research program 2011-2012,
\lq\lq Joint Usage/Research Center for Interdisciplinary Large-scale Information Infrastructures''
in Japan,
the Grant-in-Aid for the Global COE Program "The Next Generation of
Physics, Spun from Universality and Emergence" from the Ministry of
Education, Culture, Sports, Science and Technology of Japan (MEXT),
the Grant-in-Aid for Scientific Research 
[21540304 (KH), 22540243 (KH), 23540271 (KH), 24540235 (ATO)], 
and a research grant from Hokkai-Gakuen Educational Foundation.
This work was also supported by Research Fellowships of the Japan Society 
for the Promotion of Science for Young Scientists (YM).
%

%

\end{document}